\documentstyle[epsf,aps,preprint]{revtex}

\begin{document}
\draft
\preprint{\it submitted to \jcp}

\def\klung{{\lower 2pt \hbox{$<$} \atop \raise 1pt \hbox{$ \sim$} }}
\def\grung{{\lower 2pt \hbox{$>$} \atop \raise 1pt \hbox{$ \sim$} }}

\title{Exact and Semiclassical Density Matrix of a Particle\\ 
       Moving in a Barrier Potential with Bound States}

\author{Franz Josef Weiper, Joachim Ankerhold, and Hermann Grabert}  

\address{Fakult{\"a}t f{\"u}r Physik,
        Albert--Ludwigs--Universit{\"a}t Freiburg, 
              Hermann--Herder--Stra{\ss}e 3, \\
             D-79104 Freiburg im Breisgau, Germany}
\date{October 11, 1995}

\maketitle

\begin{abstract}
We present a barrier potential with bound states that is exactly solvable
and determine the eigenfunctions and eigenvalues of the Hamiltonian.
The equilibrium density matrix of a particle moving at 
temperature $T$ in this nonlinear barrier potential field 
is determined. 
The exact density matrix is compared with the result 
of  the path integral approach in the semiclassical approximation.  
For opaque barriers the simple semiclassical approximation is found 
to be sufficient 
at high temperatures while at low temperatures the fluctuation paths
may have a caustic depending on temperature and endpoints.
Near the caustics the divergence of the simple semiclassical 
approximation of the density matrix  is removed by a nonlinear 
fluctuation potential. For opaque barriers the improved 
semiclassical approximation is again in agreement with the exact result.
In particular, bound states and the form of resonance states are
described accurately by the semiclassical approach.
\end{abstract}
\pacs{PACS numbers: 82.20.Kh, 03.65.Sq, 05.30.-d }

\narrowtext

\section{Introduction}
The semiclassical approximation to quantum mechanical problems 
is useful in a large variety of contexts.
In chemistry and physics this approach becomes more 
prevalent for many reasons.
For instance, semiclassical methods are efficient for the calculation
of highly excited states, for which direct quantum mechanical 
calculations become difficult. 
The semiclassical approach also offers conceptual insights
into the dynamics of many systems that are not easily extracted
directly from the quantum mechanical treatment\cite{4}.
Finally, physical quantities of systems with barrier potentials 
can be calculated in the semiclassical limit, that corresponds to 
a large barrier height or a large barrier width\cite{6}.
Here we study a system with a nonlinear potential 
field that has a barrier and an adjacent well.
For the specific potential field considered, the 
classical equation of motion as well as the corresponding  
Schr\"{o}\-dinger equation can be solved exactly.  Therefore, the 
model can be used to test the quality of approximations 
to which one has to resort for most realistic potential fields. 

The imaginary time path integral approach  provides a 
consistent scheme for the semiclassical approximation  
of several quantities in statistical mechanics\cite{9}. 
It has been noted already 25 years ago\cite{miller}
that classical paths methods offer facile techniques
to determine the semiclassical approximation
of the equilibrium density matrix or the partition function.
In the last few years the 
formulation of quantum statistical mechanics based on 
the centroid density, originally
suggested by Feynman, was studied extensively \cite{voth}.
Although this formulation allows for highly accurate approximations for
the partition function, the centroid density has the same physical 
interpretation as
the density matrix only in the high temperature
limit.
The simple semiclassical approximation for the density matrix 
studied in the past \cite{miller}
becomes exact at high temperature, however for coordinates 
near critical values divergencies arise when the temperature 
is lowered \cite{12}.
Well-known as the problem of caustics,
these divergencies reflect the fact that new classical paths
become available.
In the region, where these paths emerge, one has to improve
upon the simple semiclassical approximation
and evaluate non-Gaussian fluctuation integrals \cite{12}.

In this article we compare the exact equilibrium density 
matrix with the improved semiclassical approximation. 
First, in section 2  we calculate the exact equilibrium density 
matrix using analytic solutions of the Schr\"{o}dinger 
equation. The simple semiclassical approximation is 
evaluated in section 3. In section 4 we present an improved 
semiclassical approximation for the density matrix which remains valid
near the caustics. Section 5 presents semiclassical results for 
bound and resonance states.

\section{Exact Equilibrium Density Matrix}

We consider the thermal equilibrium state of a quantum system with the
asymmetric barrier potential  
\begin{equation}
V_0(X)= U_0\frac{A+B \sinh( X/L_0)}{{\cosh^2( X/L_0)}} , 
\label{eq1}
\end{equation}
where $A$ and $B$ are dimensionless coefficients, $L_0$ 
denotes the typical range of the potential and $U_0 $ 
the energy scale. The Schr{\"o}dinger equation with potential (\ref{eq1})
will be seen to be exactly solvable.
In particular, for $B=0$ we recover the symmetric Eckart
potential \cite{13a,13b}.

The antisymmetric part of the potential proportional to
$\sinh( X/L_0)/ \cosh^2( X/L_0)$ is different from the 
$\tanh(X/L_0)$ term familiar from the work by 
Rosen and Morse\cite{14}.
This leads to important qualitative differences in the 
shapes of these exactly solvable potentials.
As seen from Fig.~1, the potential (\ref{eq1}) always approaches
the same limiting value for $X\rightarrow \pm \infty$, and
for $B\neq 0$ it displays a barrier with an adjacent well.
The fact that the potential is a genuine 
scattering potential decaying exponentially
for large $|X|$ and the barrier--well form distinguish 
(\ref{eq1}) from other exactly solvable smooth potentials.

\subsection{The Schr{\"o}dinger Equation}

The Schr{\"o}dinger equation 
of a quantum mechanical particle moving in the potential
(\ref{eq1}) may be written as
\begin{equation}
\left[- \frac{1}{2}\frac{d^2}{d{x}^2}+V(x)\right]
\Psi(x)= \epsilon \Psi( x)  ,
\label{eq2}
\end{equation}
with the scaled potential
\begin{equation}
V(x)=\frac{a+b \sinh(x)}{ 2 \cosh^2(x)}.
\label{scalepot}
\end{equation}
In (\ref{eq2}) and (\ref{scalepot}), we have introduced 
the dimensionless quantities
\begin{equation}
\begin{array}{lllllll}
 x&=& X/L_0& & ~~~\epsilon  &=& m L_0^2 E/\hbar^2 ,\nonumber\\
a &=& 2 m U_o L_0^2 A/\hbar^2 , & &~~~ 
b &=& 2 m U_o L_0^2 B/\hbar^2 .
\end{array}
\label{eq3}
\end{equation}
For later purpose we also introduce the dimensionless 
inverse temperature
\begin{equation}
\beta =\hbar^2 /m L_0^2 k_{\rm B} T,
\label{eq4}
\end{equation}
where $k_{\rm B}$ is the Boltzmann constant\cite{footenote}.

With the transformation $ y=\sinh(  x) $\cite{16}
the Schr{\"o}dinger equation (\ref{eq2}) takes the form
\begin{equation}
\Psi''(y) + \frac{y}{1+y^2} \Psi'(y) +
\left[\frac{2 \epsilon}{1+y^2}-\frac{a+by}{\left( 1+y^2 
\right)^2}\right] \Psi(y) = 0   .
\label{eq5}
\end{equation}
This second order differential equation has three 
singular points
at $ i, -i$, and $ \infty $. It can be reduced to its 
canonical form with the ansatz
\begin{equation}
\Psi(y)= \left( y + i\right)^{\frac{1}{4}
+\frac{1}{4}s}\left( y - 
i\right)^{\frac{1}{4}+\frac{1}{4}s^*} F(y),
\label{eq6}
\end{equation}
where
\begin{equation}
s=\sqrt{1-4a+i 4 b}=s' +i s'',  
\label{eq7}
\end{equation}
with the real and imaginary parts given by
\begin{eqnarray}
s'&=&\left\{\frac{1}{2}
\left[(1-4a)^2+(4b)^2\right]^{\frac{1}{2}}
+\frac{1-4a}{2}\right\}^{\frac{1}{2}}\nonumber \\
s''&=&\mbox{sign}(b)\left\{\frac{1}{2}
\left[(1-4a)^2+(4b)^2\right]^{\frac{1}{2}}
-\frac{1-4a}{2}\right\}^{\frac{1}{2}}.  
\label{eq8}
\end{eqnarray}
Inserting (\ref{eq6}) into (\ref{eq5}), we obtain
with the substitution
\begin{equation}
z=\frac{1}{2}-i \frac{y}{2}   
\label{eq9}
\end{equation}
a hypergeometric differential equation for $F(z)$   
\begin{eqnarray}
z(1-z) F''(z)+[\gamma -(\alpha + \alpha^* +1) z] F'(z) 
-\alpha \alpha^* F(z) = 0.
\label{eq10}
\end{eqnarray}
Here, the coefficients $\alpha$ and $\gamma $ are given by
\begin{eqnarray}
\alpha & =& \frac{1}{2} +\frac{s'}{2} + i \sqrt{2 \epsilon} 
 \nonumber \\
\gamma & =& 1 +\frac{1}{2}s.
\label{eq11}
\end{eqnarray}

A solution of (\ref{eq10}) is given by the
hypergeometric function $_2 F_1 ( \alpha ,\alpha^* ;\gamma;z)$\cite{18}.
The corresponding solution of equation (\ref{eq5}) 
reads
\begin{equation}
\Psi(z)=\left[ i 2 z\right]^{\frac{1}{4}+\frac{1}{4}s}\left[ 
i 2 (z-1) \right]^{\frac{1}{4}+\frac{1}{4}s^*}
{_2 F_1} ( \alpha ,\alpha^* ;\gamma ;z).
\label{eq12}    
\end{equation}
This solution is defined in the complex plane for $|z| \leq
 1 $, i.e.\ for
coordinates $| x| \leq \mbox{arcosh}(2 )$. 
Beyond this region the solution is  analytically 
continued
by means of the formula \cite{18}
\begin{eqnarray}
 _2 F_1 ( \alpha ,\alpha^* ;\gamma ;z) &=&
\frac{\Gamma 
(\gamma) \Gamma(\alpha^* -\alpha)}{\Gamma (\alpha^*) 
\Gamma (\gamma -\alpha )}  \left(-z\right)^{ - \alpha}
{_2 F_1} ( \alpha ,1-\gamma+\alpha;1-\alpha^* +\alpha ;
\frac{1}{z}) \hskip1cm\nonumber \\[2mm]
& &+\frac{\Gamma (\gamma) \Gamma (\alpha -\alpha^* )}
{\Gamma (\alpha) \Gamma (\gamma - \alpha^* )}  \left(
-z\right)^{ - \alpha^*}
{_2 F_1} ( \alpha^*,1-\gamma+\alpha^* ; 1-\alpha+
\alpha^* ;\frac{1}{z}).
\label{eq13} 
\end{eqnarray}
This way we have found an exact solution of 
the Schr{\"o}dinger equation defined in the whole 
domain of $ x$ . Of course, physically acceptable 
solutions have to satisfy proper boundary conditions.

Let us first consider the case $\epsilon > 0 $, i.e.\ 
a typical scattering situation. 
Since the potential (\ref{scalepot}) decays exponentially for large $|x|$,
the wave function $\Psi( x) $ should reduce to plane 
wave states at large coordinates.
For ${ x} \rightarrow \pm\infty $ the argument $1/z$ in 
(\ref{eq13}) vanishes and one can show 
that the asymptotic behavior of the solution (\ref{eq12}) 
is indeed of the form
\begin{equation}
\Psi({ x} \rightarrow \pm \infty) = A_{\pm} \exp\left[- 
i \sqrt{2\epsilon} (\pm { x})
 \right] + B_{\pm} \exp\left[i \sqrt{2\epsilon} (\pm { x}) 
\right], 
\label{eq14}
\end{equation}
with energy dependent amplitudes
\begin{eqnarray}
 A_{\pm}  = \frac{\Gamma (\gamma) \Gamma (\alpha^* -
\alpha)}{\Gamma (\alpha^*) \Gamma (\gamma -\alpha )}  
\exp \left\{
\frac{1+s'}{2}\left[\ln(2)\mp i \frac{\pi}{2}\right] 
 \pm\left[\frac{\pi}{2}\pm2 i \ln(2)\right]\sqrt{2 \epsilon}
-\left(\frac{\pi}{2}\mp \frac{\pi}{2}\right)s''\right\}
\label{eq15}
\end{eqnarray}
and
\begin{eqnarray}
B_{\pm}  = \frac{\Gamma (\gamma) \Gamma (\alpha -\alpha^* )}
{\Gamma (\alpha) \Gamma (\gamma - \alpha^* )}
\exp \left\{
\frac{1+s'}{2}\left[\ln(2)\mp i \frac{\pi}{2}\right]
\mp\left[\frac{\pi}{2}\pm 2 i \ln(2)\right]\sqrt{2 \epsilon}
-\left(\frac{\pi}{2}\mp \frac{\pi}{2}\right)s''\right\}
.\label{eq16}
\end{eqnarray}
Accordingly, the complex conjugate solution $\Psi^*({ x}) $
behaves asymptotically as 
\begin{equation}
\Psi^*({ x} \rightarrow \pm \infty) = A_{\pm}^* \exp
\left(\pm i \sqrt{2 \epsilon}  x
 \right) + B_{\pm}^* \exp\left(\mp i \sqrt{2 \epsilon} 
 x \right),
\label{eq17}
\end{equation}

We superimpose these two solutions to  
wave functions 
\begin{equation}
\Phi_\pm({ x}) =  u_\pm \Psi({ x}) +v_\pm \Psi^*({ x}) 
\label{eq17a}
\end{equation} 
with the asymptotic behavior 
\begin{equation}
\Phi_\pm({ x} 
\rightarrow \pm \infty)=
\exp \left(\pm i \sqrt{2 \epsilon}{ x}\right),
\label{eq17b}
\end{equation} 
describing outgoing scattering waves.
The coefficients $u_\pm$ and $v_\pm$ are found to read
\begin{eqnarray}
u_\pm &=& \frac{B_\pm^*}{|B_\pm|^2-|A_\pm|^2} \nonumber \\[2mm]
v_\pm &=& -\frac{A_\pm}{|B_\pm|^2-|A_\pm|^2}.  \label{eq18}
\end{eqnarray}
The asymptotic behavior of $\Phi_\pm( x)$ for $ x
\rightarrow \mp\infty $ is then given by
\begin{equation}
\Phi_\pm( x \rightarrow \mp \infty) = \frac{1}
{t_\pm(\epsilon)} \exp\left(\pm i \sqrt{2 \epsilon}{ x} \right)
+\frac{r_\pm(2 \epsilon)}{t_\pm(\epsilon)} 
\exp\left(\mp i \sqrt{2 \epsilon}{ x} \right),
\label{eq19}
\end{equation}
where the transmission and reflection amplitudes 
$t_\pm(\epsilon)$ and $r_\pm(\epsilon)$ are given by
\begin{eqnarray}
t_\pm(\epsilon)&=&\frac{|B_\pm|^2-|A_\pm|^2}{B_\pm^* A_\mp -B_\mp^* A_\pm} 
\nonumber \\[2mm]
r_\pm(\epsilon)&=&\frac{B_\pm^* B_\mp -A_\mp^* A_\pm}{B_\pm^* A_\mp -B_\mp^* 
A_\pm} . \label{eq20} 
\end{eqnarray}
The amplitudes $t_+$ and $r_+$ can be expressed in terms of
$r_-$ and $t_-$ by
\begin{eqnarray}
t_+&=&t_-\nonumber\\
r_+&=&- r_-^*t_-/t_-^*.
\label{trrel}
\end{eqnarray} 
Clearly, $t_\pm(\epsilon)\Phi_\pm({ x})$ describe the 
stationary scattering of particles incident
from the far left and right, respectively, 
with energy $\epsilon $ and unit amplitude.
Inserting the explicit result for the amplitude factors
$A_\pm $ and $B_\pm $, the transmission probability 
$T(\epsilon)=|t_\pm(\epsilon)|^2$ is found to read 
\begin{equation}
T(\epsilon) = \exp\left(\pi s''\right)
 \frac{\sinh^2\left(\pi \sqrt{2 \epsilon}
 +\frac{1}{2}\ln\left\{ \frac{\cosh\left[\pi \left(\sqrt{2 \epsilon}-s''/2\right)\right]}
{\cosh\left[\pi \left(\sqrt{2 \epsilon}+s''/2
\right)\right]}
\right\}\right)}
{\sinh^2 \left(\pi \sqrt{2 \epsilon}\right) +
\cos^2 \left(\frac{\pi}{2}s'\right)}.  
\label{eq21}
\end{equation}
For the symmetric case, $b=0$, this result  
agrees with the known transmission probability
for the Eckart potential\cite{13a,13b}.

Fig.~2 shows $T$ as a function of $b$ for fixed 
$a$ and various energies.
Note that for small energies the transmission probability
can be raised by increasing
the asymmetry parameter $b$.
While with increasing $b$ the barrier height increases 
too, the width of the barrier decreases
leading to a maximum of the transition probability 
at finite b.

For $\epsilon < 0$ the solutions of the Schr{\"o}dinger 
equation are not bound, except for the case when $\epsilon$
coincides with an energy eigenvalue.
To calculate the energy eigenvalues of the 
potential we seek for poles of the 
$S$-matrix at negative energies, or poles of the 
transmission amplitude, respectively.
This way we get for the bound state spectrum
\begin{equation}
\epsilon_n = - \frac{1}{8}\left[-(2n+1)+s'\right]^{2},
\label{eq22}
\end{equation}
in which $n=0,1,\ldots,n_{\rm max}$ where 
$n_{\rm max}$ is the largest integer with
$2n +1< s'$. From (\ref{eq2}) and (\ref{eq10}) the 
corresponding bound states are obtained as
\begin{equation}
\psi_n(z)=\frac{1}{\sqrt{N_n}}\left[ i 2 z\right]^{\frac{1}{4}-\frac{1}{4}s}\left[ 
i 2 (z-1) \right]^{\frac{1}{4}-\frac{1}{4}s^*}
{_2 F_1}\left(1+n-s' ,-n ;1-\frac{s}{2} ;z\right),
\label{eq22a}
\end{equation} 
where $z$ is given by (\ref{eq9}) and $N_n$ is a normalization constant. 
Note that the second coefficient of the hypergeometric function is 
a negative integer. Therefore the hypergeometric function
is a polynomial of $z$ of order $n$. 
Of course, $\psi_n$ is a complex function but one can show that the 
real and the imaginary parts of $\psi_n$ differ only by a constant.
Hence, the eigenvalue $\epsilon_n $ is not 
degenerate.
Clearly bound states only exist for $s' \geq 1$.

In the symmetric case $b=0$ bound states only occur for
$a < 0$ and the spectrum has the 
well known form \cite{13b}
\begin{equation}
\epsilon_n = - \frac{1}{8}\left\{-(2n+1)+\left\{1+4a\right\}^{\frac{1}{2}}
\right\}^{2}.
\label{eq23}
\end{equation}
For $a=0$ the potential (\ref{eq1}) becomes 
antisymmetric and the energy eigenvalues of the 
bound states are  
\begin{equation}
\epsilon_n = - \frac{1}{8} \left(
                                  -(2n+1)+\left\{ \frac{1}{2}+\frac{1}{2}
\left[1+(4b)^2\right]^\frac{1}{2}\right\}^{\frac{1}{2}}
\right)^{2}.
\label{eq24}
\end{equation}
In both of these cases there exists at least one bound state 
for arbitrary $a$ and 
$b$, respectively. If $  a > 0$ and $b \neq 0$ the 
condition for having at least one bound state is 
$b \geq \sqrt{a}$.

Fig.~3 shows the transmission probability 
$T$ as a function
of $a \leq 0$ for fixed energy and various 
values of the asymmetry
parameter $b$. The transmission probability for $b=0$
has maxima exactly at those points 
where new bound states are formed, i.e.\ where
$\epsilon_{n_{max}}$ 
is equal to zero. Then, for all energies the transmission 
probability reaches 1 and an incoming wave function 
is completely transmitted.
We note that a similar behavior of the transmission 
probability is also found for $b \neq 0$ and $a \geq 0$, however,
in general the position and the height of the maxima depend
on the values of $\epsilon $ and $b$.
For instance, for $\epsilon \rightarrow 0$ 
the transmission probability has 
a maximum for $a=1$ at $b=1$ (Fig.2) where
$n_{max}=0$.
Further maxima occur for $s'=2 n+1$, i.e.\
for $a=1$, at those values of b where
\begin{equation}
b=\frac{2n+1}{2}\sqrt{(2n+1)^2 +3}.
\label{eq25}
\end{equation}
The transmission probability for 
$s'=2n+1$ reads 
in the limit $\epsilon \rightarrow 0$ 
\begin{equation}
\left.T(0)\right|_{s'=2n+1}=1/\cosh^2(\pi s''/2).
\label{eq26}
\end{equation}

\subsection{The Equilibrium Density Matrix}

We assume now that the energy of particles incident 
from the far left and right is thermally distributed. 
If we use the dimensionless formulation introduced in (\ref{eq3}) 
and (\ref{eq4}), 
the state of the system is then described by the 
equilibrium density matrix at inverse temperature 
$\beta$ which has the coordinate 
representation
\begin{equation}
\rho_{\beta}(x, x^\prime)  = \frac{1}{Z}
\langle x \mid \exp (- \beta \hat{H}) \mid x^\prime \rangle   
\label{eq27}
\end{equation}
where $Z$ is an appropriate normalization constant and 
$\hat H$ the Hamilton operator from (\ref{eq2}). 
Using a complete set $\{\Phi_{\epsilon}\}$ of orthonormal 
eigenfunctions of the Hamiltonian with 
\begin{equation}
\int\limits_{0}^{\infty} d\epsilon \;\Phi_{\epsilon}^*(x)\,	
\Phi_{\epsilon}(x') + 
\sum_{n=0}^{N_{max}} \Phi_{n}^*(x)\,\Phi_{n}(x')=
\delta (x-x')
\label{eq28} 
\end{equation}
 one gets 
\begin{eqnarray}
\rho_\beta(x,x^\prime)   =  \frac{1}{Z} 
\left[
\int\limits_{0}^{\infty} d\epsilon \;\exp(-\beta \epsilon)
\Phi_{\epsilon}^*(x)\,\Phi_{\epsilon}(x') 
 +  \sum_{n=0}^{n_{max}}\exp(-\beta \epsilon_n)
\Phi_{n}^*(x)\,\Phi_{n}(x')\right] \;.
\label{eq29}
\end{eqnarray}

Hence, the density matrix at given inverse temperature $\beta $
as a function of $x$ and $x'$ is reduced to a sum over 
the discrete spectrum and an integration
about the continuous spectrum which both has to be done numerically.

The bound states in (\ref{eq29}) are given by (\ref{eq22a}).
The scattering states $\Phi_\epsilon$ are constructed appropriately
from the solutions $\Psi (x)$ in (\ref{eq12}).
According to (\ref{eq17a})
we have for given energy two independent scattering
states of the system with the asymptotic behavior given by 
(\ref{eq17b}) and (\ref{eq19}).
Hence, $t\Phi_+/\sqrt{2\pi}$ and $t\Phi_-/\sqrt{2\pi}$ 
form an orthonormal set of wave functions for $\epsilon > 0$. 
Thus, according to (\ref{eq29}), the diagonal part of the density 
matrix, i.e.\ the position
distribution $P(x)=\rho_\beta( x, x)$, 
reads 
\begin{eqnarray}
P( x) = \frac{1}{Z} \left[ 
\int_{0}^\infty 
\frac{ d k}{2 \pi} 
\exp(-\beta k^2/2 ) 
T(k)
\bigg(  |\Phi_+(k, x)|^2 +|\Phi_-(k, x)|^2  \bigg) 
+\sum_{n=0}^{n_{max}} \exp(-\beta \epsilon_n)
\frac{|\psi_n(x)|^2}{N_n} 
\right],
\label{eq33}
\end{eqnarray}
where we have introduced the dimensionless momentum $k$ by
\begin{equation}
\epsilon = \frac{1}{2}k^2.
\label{eqk}
\end{equation}
Performing the integration (\ref{eq33}) for various values 
of $a$ and $b$ one sees that for large values of 
$a $ and $b$, corresponding to potential fields with a 
large height or depth, the numerics becomes rather nontrivial because
of the strongly oscillating hypergeometric functions.
The potential (\ref{scalepot}) has extrema 
\begin{equation}
V^\pm=\frac{1}{4}\left(a\pm \sqrt{a^2+b^2}\right)
\label{vpm}
\end{equation} 
at
\begin{equation}
x^\pm=\mbox{ arsinh}\left[-\frac{a}{b}\pm 
\sqrt{\left(\frac{a}{b}\right)^2+1}\right].
\label{xpm}
\end{equation}
For increasing $a$ and $b$ the absolute values of $V^\pm$ become
large. 
In dimensional units this corresponds to the fact that the extrema
of the potential become large compared to the quantum mechanical
energy scales of the system.
The latter are given by the ground state energies of harmonic oscillators 
with frequencies proportional to the curvature of the potential
in the well and at the barrier top, respectively.
Now, for large $V^{\pm}$ we have the typical situation where a 
semiclassical approximation should be valid.
Thereby, the semiclassical expansion in powers of $\hbar $
in dimensional units corresponds to an expansion in powers of 
$1/\sqrt{2V^\pm}$ in the dimensionless formulation introduced above.
We will show in the following that the usual 
WKB--approximation is indeed sufficient except 
for a small parameter range where the simple 
semiclassical approximation has to be improved 
upon. In particular, this improved semiclassical result 
can be handled numerically very simply and 
provides an excellent
approximation for the exact density matrix in the 
region where the numerical evaluation
of Eq.\ (\ref{eq33}) becomes difficult.

\section{Semiclassical Density Matrix}
\subsection{Path Integral and Semiclassical Approximation}

The dimensionless coordinate representation of the equilibrium 
density matrix  of a quantum particle  moving in the potential
 $V(x)$  may be written as an imaginary
time path integral\cite{9}
\begin{equation}
\rho_\beta(x,x^\prime) = \int \! {\cal D}[x] \; 
{ e}^{- S[x]}  . 
\label{eq36}
\end{equation}
Here, the functional integral is over all paths $x(\tau)$, $0 
\leq \tau \leq \beta $ with $x(0) = x$ and $x(\beta) = 
x^\prime$. Each path is weighted by its Euclidean action
\begin{equation}
S[x] = \int\limits_{0}^{\beta}\!\!  d\tau\, 
\left[ \frac{1}{2} \dot{x}^2 + V(x) \right] .  
\label{eq37}
\end{equation}

To evaluate the path integral in a semiclassical expansion 
we first determine the maximum of the 
weighting factor, that is the minimum of $S[x]$. 
This is given by the classical action $S[x_{ cl}]$, 
where $x_{ cl}$ is the classical path solving the 
classical equation of motion following from Hamilton's 
principle $\delta S[x]=0$. An arbitrary path in 
(\ref{eq36}) reads
\begin{equation}
x(\tau) = x_{ cl}(\tau)  + y(\tau)  ,
\label{eq38}
\end{equation}
where $y(\tau)$ describes the quantum fluctuations 
about the classical path. The fluctuations have to 
fulfill periodic boundary conditions $y(0)=y(\beta)=0$. 
Using (\ref{eq38}) the full action is then 
expanded around its minimum.
This way the dominant part is separated off 
and one is left with a functional 
integral over periodic paths. In the simple 
semiclassical or WKB--approximation, the expansion of the full action
is truncated after the second order term leading to an 
exactly solvable Gaussian path integral\cite{9}. 

If there exists a set $\{x_{ cl}^\alpha \}$ of classical 
trajectories in $V(x)$, this procedure must be 
performed for each $x_{ cl}^\alpha$, and all 
contributions are summed to yield the semiclassical 
density matrix
\begin{equation}
\rho_{\beta}(x,x^\prime) = \sum_{\alpha}\, 
\frac{1}{\sqrt{J_{\alpha}}}\, 
{ e}^{- S[x_{ cl}^\alpha]}  ,  
\label{eq41}
\end{equation}
where $J_{\alpha}= { det}\{\delta^2 S[x]/\delta x(\tau_1) 
\delta x(\tau_2)|_{x=x_{ cl}^\alpha}\}$ is the determinant 
describing the Gaussian integral over the quantum fluctuations 
\cite{9}.
 $J_\alpha$ is given by the product of eigenvalues 
$\{\Lambda_n\}$ of the second order variational operator  
$\delta^2 S[x]/\delta x(\tau_1) 
\delta x(\tau_2)|_{x=x_{ cl}^\alpha}$ as
\begin{equation}
J_\alpha = 2 \pi  \beta \; 
\prod_{n}\, \left( N \:\Lambda_n^\alpha \right) 
\label{eq42}
\end{equation}
where $N$ is an appropriate normalization constant.      
As long as the second order variational operator  
is positive definite, i.e. $\Lambda_n > 0$ for all $n$, 
the Gaussian approximation gives the leading order 
fluctuation term.  But a 
problem arises if one of the eigenvalues $\Lambda_n$ 
tends to zero, e.g. as the temperature is lowered.  
Then, the quantum fluctuations of this mode become 
arbitrarily large and the simple semiclassical 
approximation breaks down. Generally, the vanishing 
of an eigenvalue $\Lambda_n$ defines a point where 
new minimal action paths in the potential $V(x)$ 
become possible. This is  well--known as the 
problem of caustics. 
For our purposes we also use  another 
representation of  $J_\alpha$   equivalent to  
 (\ref{eq42}). One finds\cite{9}
\begin{equation}
J_\alpha =  2 \pi \;  
\dot{x}_{ cl}^\alpha (\beta) \, 
\dot{x}_{ cl}^\alpha (0) 
\int\limits_{0}^{ \beta} \, 
\frac{1}{\left[\dot{x}_{ cl}^\alpha(\tau) \right] ^2} \, 
d\tau  \label{eq43}
\end{equation}
where $\dot{x}$ denotes the time derivative 
$ dx/d\tau$. This way the semiclassical 
approximation is completely determined by 
the classical paths.

\subsection{Classical Paths}
In the following we first investigate the classical motion 
in the inverted potential $-V(x)$.
Afterwards, the corresponding classical action and 
the contribution of the quantum fluctuations about the classical 
path are determined. 

The Euclidean energy of the system reads
\begin{equation}
\epsilon= -\frac{1}{2}\, \dot{x}^2 + \frac{a+b 
\sinh( x)}{2 \cosh^2( x)}  .
\label{eq46}
\end{equation}
The solution of  
the classical equation of motion for fixed 
$\epsilon$ is given  by
\begin{equation}
x(\tau) = \mbox{arsinh}\left\{
\frac{b}{4 \epsilon}+ 
\sqrt{\frac{b^2}{ 16 \epsilon^2}+
\frac{a}{2\epsilon}-1} \, 
\sin\left[ \sqrt{2\epsilon}\,( \tau + \beta_0) 
\right] \right\},  
\label{eq48}
\end{equation}
where $\beta_0$ is an integration constant. 
The parameters $\epsilon$ 
and $\beta_0$ of the solution are determined by 
the boundary conditions 
$x(0)= x$, $x(\beta)= x'$ 
leading to
\begin{eqnarray}
\sinh(x) & = & \frac{b}{4 \epsilon}+ 
\sqrt{\frac{b^2}{16 {\epsilon}^2}+\frac{a}{{2\epsilon}}-1}
\, \sin[\sqrt{2\epsilon} \beta_0] \nonumber \\
\sinh(x') & = &  \frac{b}{4 \epsilon}+
\sqrt{\frac{b^2}{16 \epsilon^2}+\frac{a}{2\epsilon }-1}
\, \sin\left[\sqrt{2 \epsilon} (\beta + \beta_0)\right] .
\label{eq49}
\end{eqnarray}
Due to the sine function in  (\ref{eq49}) one expects 
that there may be more than one trajectory connecting the 
endpoints for a given inverse temperature or  ``time'' $\beta $.
The maximum amplitude $x_m$ of these path is then determined by 
$V(x_m)=\epsilon$.
Note that $\epsilon $ may be negative.
Therefore $\beta_0$ is in general an imaginary phase and 
due to the relation $\sin(i x)=i \sinh(x)$ the oscillatory 
solution may change into a part of an unbounded motion.

\subsubsection{Symmetric Barrier Potential \protect{$(b=0)$}}

Let us first consider the motion of a classical particle in the 
inverted symmetric barrier potential ($a>0, b=0$)\cite{19}.
The boundary conditions (\ref{eq49}) then reduce to
\begin{eqnarray}
\sinh( x)&=&\sqrt{\frac{a-{2\epsilon}}{2\epsilon}}
\sin\left[\sqrt{2\epsilon} \beta_0\right]\nonumber\\
\sinh( x')&=&\sqrt{\frac{a-{2\epsilon}}{2\epsilon}}
\sin\left[\sqrt{2\epsilon} (\beta+\beta_0)\right].
\label{eq50}
\end{eqnarray}
The simplest case is obtained for the boundary condition
${ x} ={ x }'={ x}^+ =0$ where $x^+$ is the 
coordinate (\ref{xpm}) at the barrier top.
Then, for short times, i.e.\ high temperatures, there exists 
only the constant solution $x(\tau)=0 $ of (\ref{eq48}) 
with $\epsilon=V^+=a/2$ . When the temperature is lowered 
two new solutions with $\epsilon =\pi^2/2  \beta^2$ and 
$\beta_0= 0, \beta $ arise for 
$ \beta \geq \beta_c= \pi /\sqrt{2 V^+}$. 
These solutions describe oscillations in the well of 
the inverted potential. 
It will be shown below that the constant path, 
which is stable for high temperatures, becomes unstable for 
times $\beta > \beta_c$ where the newly emerging 
trajectories with $x_m \neq 0$ are stable (cf.\ Fig.~4). 
This change of stability is typical for a Hopf bifurcation.

For $x= x' \neq 0 $ the bifurcation scenario is 
perturbed (see Fig.~4). 
For $\beta < \beta_c $ one has only one stable path 
that continuously extends 
also to the region $\beta > \beta_c$. 
For some critical value $\tilde{\beta}_c$ two new 
branches emerge describing an oscillation to the other 
side of the well in the inverted potential. 
The solution of smaller  amplitude is 
unstable while the other one is stable. 
The critical value $\tilde{\beta}_c $ 
increases with increasing $|x|=|x'|$.  
For endpoints within the harmonic range  of the potential, 
where $|x| \ll 1 $,   the critical temperature    
$\tilde{\beta}_c $ remains close to $\beta_c$. 

Further bifurcations  appear in an analog way near  
$\beta= n \beta_c$ , $ n= 1, 2, 3, \ldots $. 
For $ x=x' = 0 $ we obtain from (\ref{eq50}) 
the parameters of the two trajectories 
bifurcating at $ \beta = n \beta_c$ 
\begin{eqnarray}
b & = & 0 , \beta /n \nonumber \\
\sqrt{2 \epsilon} \beta  & = & n \pi \;\;\;\;\;\;
n=1, 2, 3, \ldots . \label{eq51}
\end{eqnarray}
The parameters of the corresponding trajectories for  
$x=x' \neq 0$ describing the perturbed 
bifurcation near $\beta = n \beta_c$ are determined by 
\begin{eqnarray}
\sin[\sqrt{2 \epsilon}(\beta + \beta_0)] 
& = & \sin[\sqrt{2\epsilon} \beta_0]
\nonumber \\
\sqrt{2\epsilon} (\beta +\beta_0) & =& 
\sqrt{2\epsilon} \beta_0 (-1)^n + n \pi  .\label{eq52}
\end{eqnarray}
For $n$ odd the newly emerging paths are periodic in 
coordinate space with $x(\beta) = x(0) $ and 
$ \dot{x}(\beta) = - \dot{x}(0)$, while for $n$ 
even they are even  periodic in phase space with 
$x(\beta)=x(0)$, $\dot{x}(\beta)= 
\dot{x}(0)$.

In the general case $x \neq x^\prime \neq 0$ 
there is again a similar bifurcation sequence as the 
temperature is lowered.  However, for high temperatures, 
i.e. for very small times $\beta$, the solution of 
 (\ref{eq50}) connecting $x$ and 
$x^\prime$  is a segment  of an  unbounded 
trajectory in the inverted potential with $\epsilon<0$.

\subsubsection{Asymmetric Barrier Potential \protect{($a,b>0$)} }

Now, we investigate the general case $a\neq b\neq 0$ where we assume
$a,b>0$.
The classical motion in the other cases $a,b<0$ and $a<0,b>0$
as well as $a>0,b<0$ is readily obtained in an analog way.

Let us consider the boundary condition ${ x}={ x}'$ in some detail.
Then (\ref{eq49}) can be transformed to read
\begin{eqnarray}
\sqrt{2\epsilon} (\beta +\beta_0) & =& 
\sqrt{2\epsilon} \beta_0 (-1)^n + n \pi ,\nonumber\\
{ x}(0)&=&{ x}.
\label{eq53}
\end{eqnarray}
For $n$ odd the phase $\beta_0$ can easily be determined from (\ref{eq53}) 
and the energy $\epsilon $ follows from
\begin{equation}
\sinh({ x})= \frac{b}{4 \epsilon} +{(-1)}^n 
\sqrt{\frac{b^2}{16 \epsilon^2}+\frac{a}{2\epsilon}-1} 
\cos\left( \sqrt{2\epsilon}\frac{\beta}{2}\right)
\label{eq54}
\end{equation}
For $n$ even the energy $\epsilon$ is given by 
$\sqrt{2\epsilon} \beta = n \pi$ and the phase $\beta_0$ is 
determined by 
\begin{equation}
\sinh\left({ x}\right)=\frac{b \beta^2}{2 n^2 \pi^2}+
\sqrt{\frac{b^2 \beta^4}{4 n^4 \pi^4}+\frac{b \beta^2}{n^2 \pi^2}-1} 
\sin\left(n \pi \frac{\beta_0}{\beta}\right).
\label{eq55}
\end{equation}

Now, let us investigate the solutions of (\ref{eq54}) and (\ref{eq55})
in detail. 
For simplicity we first discuss the motion near the extrema of the potential
 [cf.~(\ref{vpm}) and (\ref{xpm})].
For ${x}={x}^-$ the only solution of 
the equation of motion is the constant solution 
${x}(\tau)={ x}^-$ with energy $\epsilon=V^-$
for all temperatures. 
For ${x}={x}^+$ we see that for high 
temperatures there exists only one constant 
solution ${ x}(\tau)={x}^+$ with ${\epsilon}= V^+$ (see Fig.5a).
When the temperature is lowered two new solutions appear at 
${\tilde \beta}_c <  \beta_c $ describing oscillatory solutions
to the left side of the well in the inverted potential
with ${\dot{ x}}(0) < 0$.
Note that in the symmetric case ${\tilde \beta}_c$ is always larger 
than $\beta_c$. 
The solution with amplitude $x_m$ closer to $x^+$ is 
unstable while the other solution is stable.
For the stable path $|{ x}^+-x_m|$
increases with decreasing temperature. When we arrive the temperature
\begin{equation}
\beta_{ab}\left({ x}^+\right)= \frac{4}{b}\sqrt{a+b\sinh({ x}^+)},
\label{eq56}
\end{equation}
where $x_m$ reaches
\begin{equation}
x_{ab}=\mbox{arsinh}(-a/b),
\label{eq56a}
\end{equation}
the energy $\epsilon$ changes sign and the 
oscillatory solution changes into a segment of an unbounded trajectory.
For $\beta \rightarrow \infty $ the amplitude $x_m$ tends to 
${ x}^-$ (see Fig.5a).
For the unstable solution the difference $|{ x}^+-x_m|$ 
decreases with decreasing temperature until we arrive the 
critical temperature where $\beta =\beta_c$.
There the unstable solution and the constant solution meet
in a double bifurcation point.
At $\beta = \beta_c $ the constant solution becomes unstable and
a new solution with growing amplitude appears.
This stable path describes an oscillation to the 
right side in the well of the inverted potential. 
In an analogous way further bifurcations emerge with decreasing 
temperature.

Now, for ${ x}\neq { x}^+$ and ${ x}\neq {x}^-$
four cases must be distinguished.
First, for ${ x}^-<{ x}< x_{ab}$ 
the only paths connecting ${ x}$ and ${ x}'$ 
with ${ x}={ x}'$
are segments of unbounded trajectories with ${\epsilon}< 0$. 
These paths extend from high to low temperatures. 

Second, in the case ${x}>{ x}^+$ we have at high 
temperatures only one 
solution with ${\dot{x}}(0)>0$ 
that extends continuously to low temperatures (see Fig.5a). As the
temperature is decreased two new solutions 
with amplitudes $x_m<{ x}^+$ appear at a critical inverse temperature 
${\tilde \beta}_c$ describing oscillations to the left side of the well 
in the inverted potential (see Fig.5a).
One of these solutions is stable and the other one is again unstable.
With increasing ${ x}$ the critical inverse temperature
${\tilde \beta}_c$ also increases.
At $\beta =\beta_{ab}({ x})$ (\ref{eq56}) 
the stable oscillation changes to a segment of an unbounded trajectory with 
amplitude $x_m$ that reaches ${ x}^-$ if $\beta \rightarrow \infty $.
The amplitude  $x_m$ of the unstable path
moves away from $x^+$ with increasing $\beta $ until we arrive the temperature
where $|{ x}^+-x_m|={ x}-{ x}^+$.
At this point the path that is periodic in configuration space
[${ x}={ x}',{\dot{ x}}(0)=-{\dot{ x}}(\beta)$] 
becomes a path that is periodic in phase space
[${ x}={ x}',{\dot{ x}}(0)={\dot{ x}}(\beta)$] (see Fig.5a).
One can show that this new path and all further paths bifurcating
with decreasing temperature are unstable.

Third, we consider the intermediate region $x_{ab}<{ x}<{ x}^+$.
(\ref{eq54}) defines a minimal 
inverse temperature ${ \beta_c^+}$ for which 
a bifurcation occurs. The corresponding boundary value 
${\bar x}$ is denoted by $ x_c^+$.
In the symmetric case ($b=0$) we find $ \beta_c^+ = \beta_c$ and 
$ x_c^+ = { x}^+ =0$, while in the asymmetric 
case the values $ \beta_c^+ < \beta_c$ and $ x_c^+ <
{ x}^+ $ depend on the potential parameters $a$ and $b$ and
have to be calculated numerically.
Again two cases must be distinguished.
If $x_{ab}<{ x}< x_c^+$ we have 
a bifurcation scenario similar to the 
former case ${ x}>{ x}^+$.
On the other hand, in the case $ x_c^+<{ x}<{ x}^+$, 
the high temperature solution
does not extend to low temperatures continuously.
At an inverse temperature  $\beta_c^+<\beta<\beta_c $ two new 
solutions with $x_m <{ x}^+$ emerge, a  
stable and an unstable path. 
When the inverse temperature is increased
the unstable path and the high temperature solution meet at a 
bifurcation point and disappear while the new 
stable solution extends to low temperatures.
Increasing the inverse temperature further another bifurcation of 
 a stable and an unstable solution occurs, however, now with $x_m>{ x}^+$.
 
Finally, for ${ x}< { x}^-$ again two cases have to be distinguished
(see Fig.5b).
For $x$ near $x^-$ the high temperature solutions continuously extend 
to low temperatures and no bifucations occur.
As $x$ becomes smaller a bifurcation occurs for $x=x_c^-$ 
at a critical inverse
temperature $\beta_c^- >\beta_c$ that may determined from (\ref{eq54}).
This scenario remains true for all $ x < x_c^-$.
Then a stable and an unstable path bifurcate at a critical temperature.
As the temperature is decreased further the unstable path meets the high
temperature solution in a second bifurcation point while the stable paths
extends to the low temperature region (see Fig.5b)

In the general case ${ x}\neq { x}'$ the
boundary conditions (\ref{eq49}) may be evaluated in an analogous
way and similar bifurcation sequences occur as
the temperature is lowered.

\subsection{Classical Action and Fluctuation Determinant}
Fortunately, not all extremal action paths 
have to be taken into account in the semiclassical 
approximation  (\ref{eq41}) for a given 
temperature since the  corresponding classical 
actions are not always minima of $S[x]$. 
One gains from 
(\ref{eq37}) and (\ref{eq48}) for the 
action of the classical paths 
\begin{eqnarray}
{S}_{ cl}=  - \epsilon \beta 
&+&\frac{2}{\sqrt{2\epsilon}}{ Re}
 \left[ {
    \frac{f}{\sqrt{g^2-h^2}}
  \left( {
      \mbox{ arctan}\left\{
                       \frac{f \tan\left[
                           \sqrt{2\epsilon}(\beta+\beta_0)/2
                           \right] +h}{\sqrt{g^2-h^2}}
                 \right\}  }
  \right. }
\right.
\nonumber\\
&&\left. { 
    \left. {
\hskip3cm -\mbox{ arctan}\left\{
                       \frac{f \tan\left[
                           \sqrt{2\epsilon}\beta_0 /2
                          \right] +h}{\sqrt{g^2-h^2}}
                 \right\} }
     \right) }
   \right]
\label{eq57}
\end{eqnarray}
where $ Re $ denotes the real part.
In (\ref{eq57}) we have introduced the complex coefficients 
\begin{eqnarray}
f &=& \frac{1}{2}(b+ia)\nonumber\\
g &=& \frac{b}{4 \epsilon}+i\nonumber\\
h &=& \sqrt{\frac{b^2}{16 \epsilon^2}+\frac{a}{2\epsilon}-1},
\label{eq58}
\end{eqnarray}
and the quantities $\epsilon$ and $\beta_0 $ are determined by 
the boundary conditions (\ref{eq49}). 
In the symmetric case ($b=0$) (\ref{eq57}) reduces to
\begin{eqnarray}
{S}_{ cl} & = & -\epsilon \beta 
 + \sqrt{a} 
\left( {\arctan \left\{ 
            \sqrt{\frac{a}{2\epsilon}} \tan\left[
         \sqrt{2\epsilon}(\beta + \beta_0)
                        \right]
               \right\} } {
      - \arctan \left\{
            \sqrt{\frac{a}{2\epsilon}} \tan\left[
                                \sqrt{2\epsilon} \beta_0
                                        \right]
                  \right\} }
  \right)
\label{eq59}
\end{eqnarray}
Considering first the case ${x}={x}^\prime=
{x}^+=0$ we have  for $\beta < \beta_c$ 
only the trivial solution $\bar{x}(\sigma)=0$  with 
the action 
\begin{equation}
{S}_{ cl}[0] =  a \beta/2  . \label{eq60}
\end{equation} 
At $\beta_c=\pi/\sqrt{a} $ two new branches  
emerge with the action 
\begin{equation}
{S}_{ cl}^{(1)} = \frac{\sqrt{a}}{2} 
\left( 2\pi - \frac{\pi^2}{\sqrt{a}\beta} \right) 
\label{eq61}
\end{equation}
which for $\beta>\beta_c$ is smaller than 
${S}_{ cl} [0]$. Therefore, as mentioned above, 
we have a Hopf bifurcation  at  $\beta =\beta_c.$ 
The other paths with ${x}={x}^\prime=0 $ 
bifurcating at $\beta= n \beta_c $ have the action
\begin{equation}
{S}_{ cl}^{(n)} (\beta) = n\; 
{S}_{ cl} ^{(1)} (\frac{\beta}{n}) . \label{eq62}
\end{equation}
Hence,  the  actions  of  the classical paths  
appearing  at  bifurcation points for temperatures 
$\beta \geq 2 \beta_c$ differ from 
${S}_{ cl}^{(1)}$ by terms which are large 
compared to 1 in the semiclassical  limit $\hbar 
\rightarrow  0$, which, more precisely, corresponds 
to the limit $1/\sqrt{a} \rightarrow 0$ for the present case.
As will be seen below, also in the asymmetric case
at most two classical trajectories contribute to the 
semiclassical approximation of the equilibrium density matrix.

We now proceed to  calculate the Gaussian fluctuations 
about the extremal action paths. 
After some algebra one obtains from (\ref{eq43}) 
and (\ref{eq48}) for the determinant 
$ J$ 
\begin{eqnarray}
{ J}&=&\frac{2\pi}{\sqrt{2\epsilon}} 
            \left(
              1+\left\{
\frac{b}{4 \epsilon}+h \sin\left[
                \sqrt{2\epsilon}(\beta+\beta_0)
                            \right]
                \right\}^2
            \right)^{-1/2} 
            \left(
              1+\left\{
\frac{b}{4\epsilon}+h \sin\left[
                \sqrt{2\epsilon}\beta_0
                            \right]
                \right\}^2
           \right)^{-1/2}\nonumber\\
&\times & 
\Bigg[
   \left(
          \frac{b^2}{8\epsilon^2}+ \frac{a}{2\epsilon}
   \right)
       \sin\left[  
                \sqrt{2\epsilon}\beta
            \right] + 
     \frac{b}{2\epsilon}h 
 \left\{
    \cos\left[
              \sqrt{2\epsilon}\beta_0
        \right]- 
    \cos\left[
             \sqrt{2\epsilon}(\beta+\beta_0) 
        \right]    
 \right\} \nonumber \\
&-&h^2 \sqrt{2\epsilon} \beta \cos\left[
                               \sqrt{2\epsilon}\beta_0
                               \right]
           \cos\left[
                               \sqrt{2\epsilon}(\beta+\beta_0)
                               \right] 
\Bigg],
\label{eq63}
\end{eqnarray}
with $h$ from (\ref{eq58}).
Using  (\ref{eq49}) one can show that $J$ is 
positive for the stable
classical paths but negative for the unstable one.  
Of course, for unstable
paths the  fluctuation determinant  is only defined 
as an analytical continuation, or formally by 
(\ref{eq41}) and (\ref{eq42}). 
Further, one sees that  $J$  vanishes at 
the inverse temperatures
$\beta=\tilde{\beta}_c$ where the new stable 
and unstable paths emerge. We note that (63)
is also valid for $\epsilon<0$ by virtue of the relations
$\sin(i x)=i \sinh(x)$ and $\cos(i x)=\cosh(x)$. 
For instance, for ${x}={x}'={ x}^-$, the only solution 
of (\ref{eq46}) is given by $\epsilon= V^-$ and the fluctuation 
determinant $ J$ reads
\begin{equation}
{ J}[{ x}^-]= \frac{2\pi}{\sqrt{2 |V^-|}} 
\, \sinh\left(
             \sqrt{2 |V^-|}\beta\right).
\label{eq64}
\end{equation}

In particular, when we insert  the trivial solution for 
${x}={x}^\prime={ x}^+$ into (\ref{eq49}) one gets
$\epsilon=V^+$ and (\ref{eq63}) reduces to
\begin{equation}
{J}[{ x}^+] =  \frac{2\pi}{\sqrt{2 V^+}}\,    
\sin\left(
                               \sqrt{2 V^+}\beta \right)
\label{eq65}
\end{equation}
which  vanishes for $\beta = \beta_c$ and is negative for 
$\beta>\beta_c$.

Let us now discuss the main features of the fluctuation 
determinant in the symmetric case ($b=0$).
Qualitatively similar reasoning applies
to the general case.
For the stable branches with ${ x}={x}^+=0$ 
bifurcating at  
$\beta = \beta_c$ one finds for  $\beta > 
\beta_c$
\begin{equation}
{J}[{ x}^+] = 2 \, 
\frac{2 V^+ \beta^3- \pi^2 \beta }{\pi} > 0 . 
\label{eq66}
\end{equation}
This demonstrates  the change of stability near 
$\beta= \beta_c$ and shows that the simple 
semiclassical approximation becomes insufficient for 
temperatures close to $\beta_c=\pi /\sqrt{2 V^+}$, 
i.e.~$T_c=\sqrt{2 V^+}\hbar^2/
\pi k_{ B}m L_0^2$ in dimensional units.

At inverse temperature $\beta =2 \beta_c$ and 
for boundary conditions ${x}={x}^+$ again two 
new classical paths bifurcate from the unstable trivial 
solution ${x}(\sigma)=0$. However, using 
Eqs.\ (\ref{eq50}) and  (\ref{eq63}), one sees 
that the corresponding determinant ${J}$ is negative, 
while ${J}[{ x}^+]$ becomes positive for 
$2 \pi <\sqrt{2 V^+} \beta<3 \pi$. In function space 
this means that the paths newly emerging  at 
$\beta =2 \beta_c$ are unstable in one 
direction corresponding to one  negative eigenvalue 
of the second order variational operator, see 
 (\ref{eq42}), while the trivial solution is 
unstable in two directions with two negative eigenvalues 
leading to a positive fluctuation determinant. 
More generally, one can show  that all classical paths  
newly emerging for inverse temperatures 
$\beta \geq 2 \pi /\sqrt{2 V^+}$ are unstable due to 
the fact that successively the eigenvalues 
$\Lambda_n$ of the second order variational 
operator change from positive to negative 
values as the temperature is lowered. The 
only stable paths for inverse temperatures 
$\beta> \beta_c$ are those bifurcating 
first near $\beta=\beta_c=\pi/\sqrt{2 V^+}$. These 
paths are therefore the only ones contributing 
to  the functional integral for lower 
temperatures in the semiclassical limit .

For temperatures close to $T_c$ a change of 
stability occurs, as described above.    
The minimal action paths are then not well 
separated in function space and the semiclassical 
approximation according to  (\ref{eq41}) 
breaks down. This problem will be addressed in section 4.

\subsection{Semiclassical Density Matrix}

Now, for high temperatures, $\beta< \beta_c^+ $ 
and given endpoints ${x},{x}^\prime$ there is 
only one classical path with amplitude $x_m$ which is obtained from 
 (\ref{eq48}) by choosing $\epsilon$ and $\beta_0$ 
according to  (\ref{eq49}). In practice, $\epsilon$ and 
$\beta_0$ have to be determined numerically.  
By virtue of 
Eqs.\ (\ref{eq57}) and (\ref{eq63}) the 
semiclassical density matrix for high temperatures, 
$\beta < \beta_c^+$, is then given by
 \begin{equation}
 \rho_{\beta}({x},{x}^\prime) = \frac{1}{Z} \, 
\frac{1}{\sqrt{{J}(x_m)}} \; 
{ e}^{- {S}(x_m) }  .
 \label{eq67}
 \end{equation}

To illustrate this result we have depicted in 
Fig.\ref{fig:hightc} the diagonal part of the density matrix 
$P({x})= \rho_\beta({x},{x})$ for the symmetric 
potential with  $b=0$ in
the semiclassical approximation   (\ref{eq67}) 
and the exact result  (\ref{eq33}) at inverse 
temperature $\beta=\beta_c /2$ and for two values of $a$. 
Note that even for a small $a=4$, the 
semiclassical approximation does rather well. 
The probability density $P(x)$ is normalized 
by the condition $\lim_{x\rightarrow\pm\infty} P(x)=1$

For $\beta \grung  \beta_c^+$ new classical 
paths may emerge, first for coordinates near the critical 
coordinate $ x_c^+$. Then,  one unstable path 
and two stable ones exist, determined by the 
solutions $\epsilon$ and 
$\beta_0$ of  (\ref{eq49}). 
Apart from a narrow region about the critical inverse 
temperature where the bifurcation occurs,
these trajectories are well 
separated in function
space and the sum in (\ref{eq41}) contains 
the contribution of 
the two stable paths. The density matrix  in the semiclassical 
approximation then reads
 \begin{equation}
 \rho_{\beta}({x},{x}^\prime) = \frac{1}{Z}\, 
\left[ \frac{1}{\sqrt{ {J}(x_{m1})}}\; 
{ e}^{- {S}(x_{m1})} +  
\frac{1}{\sqrt{{J}(x_{m2})}}\; 
{ e}^{- {S}(x_{m2})} \right]  , 
\label{eq68}
\end{equation}
where we have denoted the amplitudes of the two stable paths with 
$x_{m1}$ and $x_{m2}$.
When ${S}(x_{m1})$ exceeds  
${S}(x_{m2})$ by terms that are 
large compared to 1, it is inconsistent to take  
both contributions in  (\ref{eq68}) 
into account. In section 4 we will show that this is 
the case for all coordinates outside a narrow region 
around the barrier top and around the well region at $x_c^-$. 
Hence, mostly the functional integral is dominated by the 
contribution of one path 
and the semiclassical density matrix reduces 
to  (\ref{eq67}).

In Fig.\ref{fig:lowtc} the semiclassical approximation for $P({x})$ 
calculated from  (\ref{eq67}) 
is depicted together with the exact result (\ref{eq33}) at inverse 
temperature $\beta = 2 \beta_c$ for the potential (3) 
with $a=b=4$. 
From (\ref{eq23}) we see that there is only one bound state.
Hence, in the semiclassical approximation deviations become apparent in 
the well region, while for the barrier region the semiclassical result gives
an excellent approximation to the exact result. 
The semiclassical approximation in the well region becomes better if 
either $a $ or $b$ are increased corresponding to more bound states
in the well. Only for coordinates in 
the vicinity of the barrier top the semiclassical 
approximation is given by the sum  (\ref{eq68}), 
which matches for larger ${x}$ onto the simpler 
formula  (\ref{eq67}).

As we have seen  above, the simple semiclassical approximation 
breaks down near those inverse temperatures $\beta $ where bifurcations 
of the classical paths occur.
As we will see in the next section
this problem becomes relevant for coordinates in a narrow region
around the barrier top, near the critical coordinate $x_c^+$, 
and in the well region, near the critical coordinate $x_c^-$.
Only there, the simple semiclassical determination has to be improved
due to the fact that
the relevant classical paths are not well separated in function
space. Furthermore, this analysis also gives the 
precise conditions for the validity of the formulae 
 (\ref{eq67}) and (\ref{eq68}).

\section{Semiclassical Density Matrix Near Caustics}

The specific results presented in Figs.~(6) and (7)
are for temperatures well above or below the critical temperatures
$\beta_c^+$ and $\beta_c^-$,
respectively, where the simple semiclassical results
(66) and (67) are valid. 
A more refined treatment is required near the critical 
temperatures.
In \cite{12} we have shown how 
to determine the semiclassical
density matrix in the critical region near the barrier top. 
Here, we shall apply these findings to the present problem.
 
Near the barrier top the potential (\ref{eq3}) with $a>0$ may be expanded as 
 \begin{eqnarray}
 V(\xi) = V^+\Bigg[1 -\xi^2 +\frac{2}{3} a_3\xi^3+\frac{1}{2}a_4 \xi^4 
+{\cal O}\left(\xi^5\right)\Bigg] ,
 \label{eq70}
 \end{eqnarray}
where $\xi = { x}-{ x}^+$ and the potential parameters
are given by
\begin{eqnarray}
a_3 &=&\frac{3}{2}\tanh\left({ x}^+\right)\nonumber\\
a_4 &=&\frac{4}{3}-\frac{5}{2}\tanh^2\left({ x}^+\right).
\label{eq70a}
\end{eqnarray}
Of course, for  coordinates   near the barrier top   
$|\xi| \ll 1 $, 
and the anharmonic terms in Eq.\ (\ref{eq70}) are much smaller than  
the parabolic one. However, for $\beta \approx { \beta_c^+}$  
they will be found to be crucial for the fluctuation integral even 
in the semiclassical limit. 
 
To see this more clearly,  we use the  representation
(\ref{eq42}) of the determinant ${J}$  as a  product of 
eigenvalues of the second order variational operator.
Representing  the scaled quantum fluctuations in the form
\begin{equation}
{y}({\tau}) =   
 \frac{1}{{\sqrt{2V^+}}} \sum_{n=1}^{\infty}\, Y_n\, 
\sin\left(\frac{\pi n}{{\beta}}{\tau} \right) ,  
\label{eq74}
\end{equation}
we have shown in \cite{12} that for high temperatures 
the second order variational operator is 
transformed to a diagonal matrix with eigenvalues 
\begin{equation}
\lambda_n  =  \left(\frac{\pi n}{\sqrt{2V^+}\beta}\right)^2 -1 .
\;\;\;n=1,2,3,\ldots 
\label{eq75}
 \end{equation}
Then, the  path integral over the fluctuation modes  
reduces to a product of independent Gaussian integrals.  
By virtue of  (\ref{eq42}) and (\ref{eq75})  
one obtains our earlier result (\ref{eq65}) 
for ${J}$ 
by inserting the trivial solution.

Now, it is obvious that the Gaussian approximation  
becomes insufficient for inverse temperatures near $\beta_c$ where 
the eigenvalue $\lambda_1$ vanishes. As the temperature 
is lowered,  the  fluctuation mode amplitude $Y_1$ 
increases with decreasing $\lambda_1$. 
As discussed in \cite{12} the anharmonic terms 
lead to new stability and one 
gets for inverse temperatures near $\beta_c$ for the action
\begin{equation}
{S}[{x}]- {S}[{x}_{\rm cl}] = 
\frac{\beta}{4}\, \sum_{n=2}^{\infty} \, 
\lambda_n Y_n^2 + V(\xi_m,Y_1)  
\label{eq78}
\end{equation}
with the fluctuation potential
\begin{eqnarray}
V( \xi_m,Y_1) = \frac{\beta}{4} \Bigg\{ 
 \left[    \lambda_1 -\lambda_c +
  \frac{9}{4} a_4 \left( \xi_m-\xi_{m,c} \right)^2 \right] Y_1^2 
+ \frac{3 a_4 }{ \sqrt{8 V^+}} \left( \xi_m- \xi_{m,c} \right) Y_1^3 
    +\frac{3 a_4}{16 V^+} Y_1^4    \Bigg\}  
\label{eq79}.
\end{eqnarray}
Here, we have introduced the critical parameters
\begin{eqnarray}
\xi_{m,c}=-\frac{32 a_3}{27 \pi a_4}
\label{eq81}
\end{eqnarray}
and
\begin{eqnarray}
\lambda_c= \frac{9}{4}a_4 \xi_{m,c}^2 .
\label{eq80}
\end{eqnarray}
The coefficients $a_3$ and $a_4$ are given by (\ref{eq70a}).
In (\ref{eq79}) $ \xi_m $ is determined by the cubic equation
\begin{equation}
\frac{3}{4} a_4 \xi_m^3 
+ \frac{8}{3 \pi} a_3 \xi_m^2
+ \lambda_1 \xi_m
 = \frac{2 \pi}{\beta^2}\, \left(\xi+\xi'\right) .  
\label{eq82}
\end{equation}
This equation has one or three real solutions corresponding 
to the one or three classical paths. In fact, for given 
endpoints near the barrier top the amplitudes $x_m$  of the 
classical paths determined by (\ref{eq49}) and $V(x_m)=\epsilon$ 
are given by the roots of Eq.\ (\ref{eq82}) according to 
$x_m =\xi_m +x^+$.

From Eq.\ (\ref{eq79}) one sees that the coefficient
\begin{equation}
\Lambda_1(\xi_m) = \lambda_1 -\lambda_c +
      \frac{9}{4} a_4 \left( \xi_m - \xi_{m,c} \right)^2
\label{eq83}
\end{equation}
of the harmonic term may vanish, but the remaining 
terms of the fluctuation potential always constrain 
$Y_1$ to fluctuation amplitudes at most of order 
$ (V^+)^{3/8}$.
Introducing the sum variable $r=(\xi + \xi ')/2$ we see 
from (\ref{eq82}) that for given $\beta$ the coefficient 
$\Lambda_1(\xi_m)$ vanishes at values $r=r_\pm $ (see Fig.\ref{r3}) where
\begin{equation}
r_\pm (\lambda_1)= r_c \left[
3 \frac{\lambda_1}{\lambda_c}-2 
\pm 2 \left(1-\frac{\lambda_1}{\lambda_c}
\right)^{3/2} \right]
\label{eq84}
\end{equation}
with
\begin{equation}
r_c=-
\frac{3 a_4 V^+{\beta}^2}{8 \pi} \xi_{m,c}^3.
\label{eq85}
\end{equation}
Note, that $r_c=x_c^+$ for $\xi=\xi '$. 
On the two curves $r_\pm(\lambda_1)$
bifurcations of classical paths with endpoints near the barrier top occur
according to the discussion in subsection 3.2.2.
This analysis gives in detail the regions in the 
$\lambda_1-r-$parameter space where one or three classical 
paths exist (see Fig.\ref{r3}). 

For fixed $\xi ,\xi '$, the potential $V( \xi_m,Y_1)$ behaves as 
follows (see Fig.\ref{flukplot}). At high temperatures the 
fluctuation potential exhibits one minimum at $Y_1=0$ 
according to one real solution of (\ref{eq82}). When the 
temperature is lowered to the value where 
$r_\pm(\lambda_1)=r $ two new extrema arise at $Y_1\neq 0$.
Then, the fluctuation potential shows two minima
and one local maximum.
In the region in coordinate--temperature space around $(r_c,\lambda_c)$
where $|\Lambda_1(\xi_m)|$ is smaller than order $(V^+)^{-1/4}$
both minima differ
by terms which are smaller than order 1 and are not separated by a
local maximum exceeding the minima by terms which are larger than
order 1.

In this region the functional integral  over the fluctuation modes now takes 
the form \cite{12}
\begin{equation}
\int {\cal D}[{y}]\; {\rm e}^{({S}-
{S}_{\rm cl})[{y}]} = \sqrt{\frac{\lambda_1 \sqrt{2V^+}}{2 
\pi\sin(\sqrt{2V^+} \beta)}} \; K(\xi_m)  
\label{eq91}
\end{equation}
with the fluctuation integral
\begin{equation}
K(\xi_m) = \sqrt{\frac{\beta}{4 \pi}} \, 
\int\limits_{-\infty}^{\infty} \! dY_1 \;{\rm e}^{- V(\xi_m,Y_1)}
 \label{eq92}
\end{equation}
describing the contribution of the marginal mode $Y_1$. 
The fluctuation integral   is finite for positive and  
negative values of $\lambda_1-\lambda_c$.

By virtue of (\ref{eq57}) and  (\ref{eq91}) 
the dimensionless density 
matrix near the top of the barrier potential (\ref{eq3}) and 
in the vicinity of $\lambda_1=\lambda_c$ takes the form
\begin{equation}
\rho_{\beta}({x},{x}^\prime)= \frac{1}{Z} \,  
\sqrt{\frac{\lambda_1 \sqrt{2V^+}}{2 \pi \sin(\sqrt{2V^+}\beta)}} 
\, K(\xi_m) \,
{\rm e}^{-{S}(x_m)} .  
\label{eq93}
\end{equation}
Usually, explicit results for the distribution can be 
calculated only numerically. After solving  (\ref{eq82}) 
for  given $\lambda_1$ and $\xi,\xi^\prime$ one has to insert 
$ \xi_m$ into the fluctuation potential  (\ref{eq79}) 
and determine the fluctuation integral (\ref{eq92}). 
All steps involve only rather simple numerics. 
Outside the critical region where $|\lambda_1-\lambda_c|$ $ \klung
$ $(V^+)^{-1/4}$ and $|\xi|,|\xi'|$ $\klung $ 
 max( $a_3 (V^+)^{-1/4}$ , $(V^+)^{-3/8} $ ) the result 
(\ref{eq93}) smoothly  matches onto the various semiclassical 
approximations specified above in  (\ref{eq67}) and 
(\ref{eq68}). 
In particular, one can show \cite{12} that for coordinates outside the
critical region, the minimum of the fluctuation potential $V(\xi_m,Y_1)$
corresponding to the globally stable path differs from the other 
minimum by terms which
are large compared to 1.
Therefore, for coordinates outside the critical region
the density matrix is given by (\ref{eq67}). 
For coordinates near the barrier top,   
 (\ref{eq67})  must be used for 
$\lambda_1- \lambda_c $~$>$~$(V^+)^{-1/4}$,  
(\ref{eq68}) for $ \lambda_1- \lambda_c $~\-$<$~\-$ (V^+)^{-1/4}$, and 
(\ref{eq93}) in between. 

In principle, the procedure to determine the fluctuation integral
(\ref{eq91}) for critical coordinates in the well region near $x_c^-$
is similar to the case studied above.
As we have seen in section 3, 
a stable and an unstable path newly emerge at the caustic
(see Fig.~5b).
These paths with endpoints near $x_c^-$ have amplitudes $x_m$
with large $|x_m-x_c^-|$. 
Hence, the trajectories explore the nonlinear range of the well region 
and the eigenfunctions of the second order variational operator
cannot be calculated analytically. 
However,  the features of the fluctuation 
potential are the same as discussed above for the caustic near $x_c^+$.
Thus the fluctuation integral can also be determined in the following way.
First, since there is one unstable path, the determinant $J$ in 
(\ref{eq63}) contains only one vanishing eigenvalue at the caustic.
Second, well beyond the caustic one has one or two stable classical paths
which are well separated in function space so that the density matrix
is given by (\ref{eq67}) and (\ref{eq68}).
Finally, the equation for the boundary values (47) expanded around the critical
coordinate $x_c^-$ and the critical temperature $\beta_c^-$
yields in leading order a cubic equation similar to (81) 
for $\xi_m=x_m-x_{m,c}$ where $x_{m,c}$ is the corresponding amplitude
at the critical point.
The critical values $x_c^-$ and $\beta_c^-$ are determined by (47)
and derivatives thereof.
Hence, similar to (78), the fluctuation potential near the critical point
may be written in the form 
\begin{equation}
V(x_{m},Y)=\frac{1}{2}{ J}(x_{m})Y^2 +\frac{1}{3}J_3Y^3+
\frac{1}{4}J_4Y^4~.
\label{eq94}
\end{equation}
Here, $J(x_{m})$ denotes the determinant of the second order
variational operator of a stable classical path with amplitude $x_m$
and $Y$ is proportional to the amplitude of the marginal fluctuation mode.
The coefficients $J_3$ and $J_4$ are determined by the corresponding
cubic equation. 
Then, the density matrix near the caustic is given by
\begin{equation}
\rho_{\beta}(x,x')=\exp\left[-{ S}(x_{m})\right]\int_\infty^\infty 
\frac{{\rm d}Y}{\sqrt{2 \pi}}
\exp[-V(x_{m},Y)],
\label{eq95}
\end{equation}
where $S(x_m)$ is the action (\ref{eq57}).
The  regions in temperature--coordinate--space
where the various  semiclassical formulas have to be used 
may be determined in a way 
analogous to the analysis near the barrier top.

Fig.~10   shows the divergent simple semiclassical result 
(\ref{eq67}), the improved formula (\ref{eq93}), 
and the exact result (\ref{eq33}) for the position 
distribution $P({x})$ at temperature $\beta=\beta_c^+$. The 
exact result and the improved formula differ by terms which 
are at most of order $1/\sqrt{2V^+}$.

The formulas (\ref{eq93}) and (\ref{eq95}) for the 
improved semiclassical approximation
combines with the formulas (66) and (67) for the usual semiclassical 
approximation to yield the semiclassical equilibrium density matrix
for all temperatures. 
In particular, the position distribution for very low temperatures
will be discussed in the next section.

\section{Semiclassical Bound States and Resonances}

From (34) one sees that for low temperatures
the density matrix is dominated by the contributions of  
bound states while the contribution of the 
extended states is exponentially suppressed.
For large $\beta $ we expect large deviation
between the semiclassical approximation (67) 
and the exact result (35) due to the fact 
that differences between the 
exact and semiclassical energy eigenvalues of the
bound states
are exponentially amplified in the corresponding Boltzmann
factors for $\beta \rightarrow \infty$.
However, we will show in the following that the semiclassical
position distribution can nevertheless be used to determine
the absolute square of the wave functions 
$|\Psi_n (x)|^2 , n=0,1..n_{max}$ of the bound states
with energies $E_n , n=0,1..n_{max}$ reasonably well.

For very low temperatures the position distribution 
function is, apart from
exponentially small corrections, dominated by the ground state.
Thus, the low temperature limit enables us to extract the absolute 
square of the ground state wave function from the shape of $P_\beta (x)$.
Additionally, from the low temperature behavior of the maximum of 
$P_\beta (x)$ in the well region we can extract
the semiclassical energy eigenvalue of the ground state $E_0$ 
by use of the formula
\begin{equation}
E_0=\frac{1}{\Delta\beta}\lim_{\beta\rightarrow \infty}\ln\left[\frac{P_\beta (x_{max})}
{ P_{ \beta+\Delta\beta }(x_{max})}\right] ,
\label{eigen}
\end{equation} 
where $x_{max}$ denotes the coordinate where $P_\beta (x)$ has its
maximum. 
The wave function $|\Psi_0(x)|^2$ is then given by
$\exp(\beta_0 E_0 ) P_{\beta_0}(x)$ where $\beta_0$ has to be choosen
sufficiently large so that this product becomes independent of
temperature \cite{note}.
This procedure can be extended also to the next exited 
state in the following way.
We substract the contribution of the ground state from the
probability distribution at a higher temperature $\beta_1$ where
$\beta_1$ has to be choosen large enough so that the reduced 
position distribution 
\begin{equation}
P_{1,\beta}(x)=P_\beta (x)-|\Psi_0(x)|^2
\exp(-\beta E_0)
\label{redp}
\end{equation}
is dominated by the first exited state.
From the reduced probability distribution we extract the first exited 
state $|\Psi_1(x)|^2$ and its energy $E_1$.
Accordingly, we get a sequence of temperatures 
$\beta_0 > \beta_1 >\beta_2 >...$ and reduced probabilities 
$P_{n,\beta}(x)$ from which we can find the semiclassical
approximation for the wave functions $|\Psi_n (x)|^2$ and the energies
$E_n$. To illustrate this, we 
depict in Fig.~11a the exact (27) and semiclassical ground
state wave function $|\Psi_0(x)|^2 $ for an antisymmetric potential
with potential parameters $a=0$ and $b=4$.
For these parameters there is only one bound state (cf.~(26))
with the semiclassical energy 
$E_0=-0.326..$, while the exact energy eigenvalue from (29) 
is $\epsilon = -0.459..$ . The temperature choosen to define
the ground state is $\beta_0 = 12$.
Note that the wave function is very well reproduced, while 
the Boltzmann factors differ largely: $\exp(\beta_0(
E_0 -\epsilon_0)) \simeq 5 $.

The semiclassical density matrix can also be used to study
scattering states, in particular resonances. 
As we have seen in the previous sections the semiclassical
approximation of the equilibrium density matrix does rather well 
even for systems that have only a few bound states except for very 
low temperatures where the amplification of errors in the 
energy eigenvalues
has to be taken into account.
From the exact solution of the equilibrium density matrix 
we know that the contribution of the scattering states
exhibits several peaks.
These resonances 
correspond to energy "eigenstates" with complex eigenvalues.
In Fig.~11b we depict the exact and semiclassical 
position distribution (upper curves) 
as well as the reduced position distribution $P_{1,\beta}(x)$ 
(\ref{redp})(lower curves) 
for the same set of parameters as in Fig.~11a at temperature $\beta =0.5$.
Since for these parameters there is only one bound state, 
$P_{1,\beta}(x)$
represents the contribution of the scattering states
to the position distribution. 
According to the single bound state in the potential, 
the distribution $P_{1,\beta}(x)$ shows one resonance peak
which is a typical quantum
mechanical effect of interference of scattering states.
The semiclassical approximation, based on purely
imaginary time mechanics, reproduces the resonance very well.

\acknowledgements
The authors would like to thank E.~Pollak for enspiring
discussions. This work was supported in part by the Deutsche
Forschungsgemeinschaft through SFB 276.

\newpage
\begin{figure}
\caption{The potential (3) for various potential
parameters $a$ and $ b$.}
\label{fig1}
\end{figure}

\begin{figure}
\caption{The transmission probability (6)
as a function of $ b$ for fixed potential parameter
$ a=1 $
and various energies $\epsilon $.}
\label{fig2}
\end{figure}

\begin{figure}
\caption{The transmission probability (6)
as a function of $ a$ at fixed energy
$\epsilon =0.015$
for various potential parameters $b $.}
\label{fig3}
\end{figure}

\begin{figure}
\caption{   
Bifurcation diagram for the amplitude $x_m$ 
of the classical path $x(\tau)$ versus the 
dimensionless inverse temperature $\beta$ for potential parameters
$a=1$, $b=0$. The solid lines 
are the stable and the dashed lines are the unstable branches 
for a pure ($x= x^\prime=0$, thick lines) and a 
perturbed ($x= x^\prime > 0$, thin lines) bifurcation.  }
\label{fig:bifur}
\end{figure} 

\begin{figure}
\caption{ 
The asymmetric bifurcation scenario for various boundary conditions
and potential parameters $a=1$, $b=5$.
a) The thick lines represents the amplitude $x_m$ for $x=x^\pm$, the
thin lines represent the amplitude $x_m$ for $x > x^+$.
The solid (dashed) lines represent the stable (unstable) paths.
b) The amplitude $x_m$ for the boundary values (from above) $x^-> x>x_c^-$, $x=x_c^-$ and
$x<x_c^-$   }
\label{fig:bif}
\end{figure}

\begin{figure}
\caption{ Position distribution $P({x})$ 
for a symmetric barrier potential ($b=0$) at inverse 
temperature with  $\beta=\beta_c /2$  for (a) $a=4$ and (b) 
$a= 25$. The solid lines represent the exact 
result  (\protect\ref{eq33}) and  the dotted 
lines the semiclassical formula  (\protect\ref{eq67}).}
\label{fig:hightc}
\end{figure}
	
\begin{figure}[t]
\caption{ 
Position distribution $P({x})$ at inverse temperature 
$\beta= 2 \beta_c$. The potential parameters are
$a=b=4$.
The solid line represents the exact result 
 (\protect\ref{eq33}), the dashed lines the 
semiclassical formula  (\protect\ref{eq67}). }
\label{fig:lowtc}
\end{figure} 	

\begin{figure}[t]
\caption{The $\lambda_1$-$r$-plane is divided
by the two curves $r_\pm (\lambda_1)$ (solid lines)
into two regions in which the cubic equation (\protect\ref{eq82})
has one or three (shaded region) solutions.
Along the line $r_e(\lambda_1)$ one has $ \xi_m= \xi_{m,c}$.}
\label{r3}
\end{figure} 

\begin{figure}[t]
\caption{The fluctuation potential (\protect\ref{eq79}) for 
\protect{$ \xi_m=\xi_{m,c}$} and inverse 
temperatures near \protect{$\beta_c$}.\hskip5cm}
\label{flukplot}
\end{figure} 

\begin{figure}[t]
\caption{ Position distribution ${P(x)}$ for a 
symmetric barrier potential with $a=25$ and $b=0$ 
at the critical inverse temperature 
$\beta_c^+ = \pi/5$. 
The solid line  represents the exact result (35), 
the dashed line the divergent semiclassical formula (66), 
and the dotted  line 
the improved semiclassical result (90).}
\label{fig:tc}
\end{figure} 
	
\begin{figure}[t]
\caption{ a) The absolute square of the
ground state wave function for an 
antisymmetric barrier potential with $a=0$ and $b=4$. 
b) The position distribution 
${P(x)}$ (upper curves) and the resonance (lower curves) for  
the same set of potential parameters
at inverse temperature $\beta=0.5$. 
The solid lines  represents the exact results, 
the dashed lines the semiclassical ones.}
\label{fig:geb}
\end{figure}

\newpage
\begin{figure}
\begin{center}
\leavevmode
\epsfysize=13cm
\epsffile{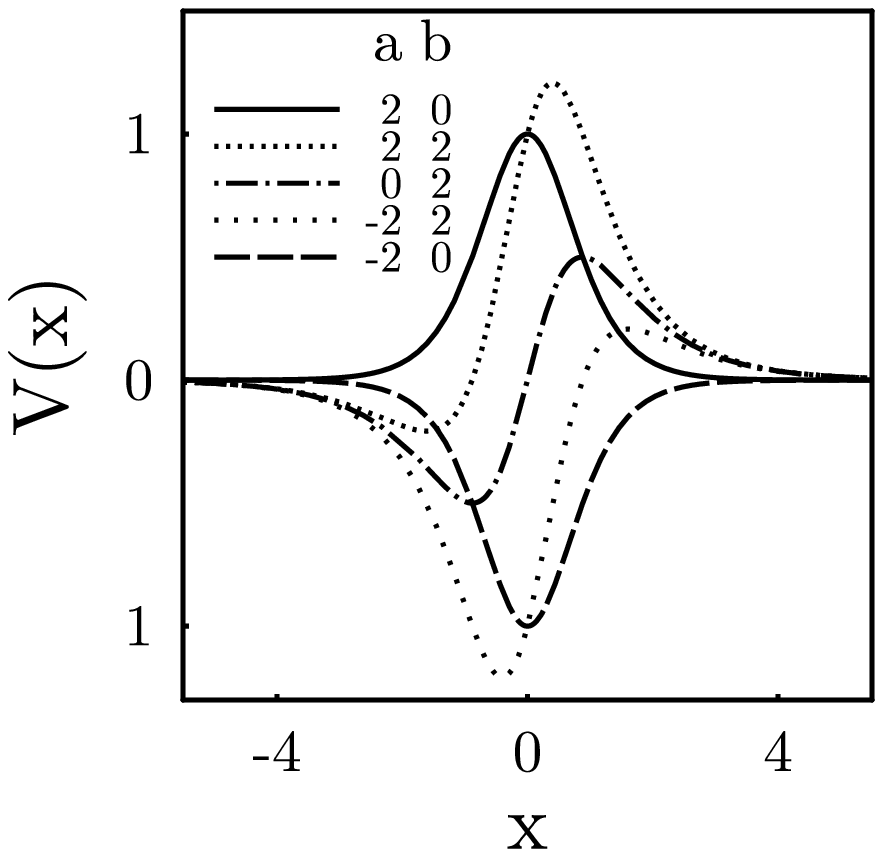}
\end{center}
\end{figure}
\vskip5cm
{\large Figure 1, F.~J.~Weiper et al.}

\newpage
\begin{figure}
\begin{center}
\leavevmode
\epsfysize=13cm
\epsffile{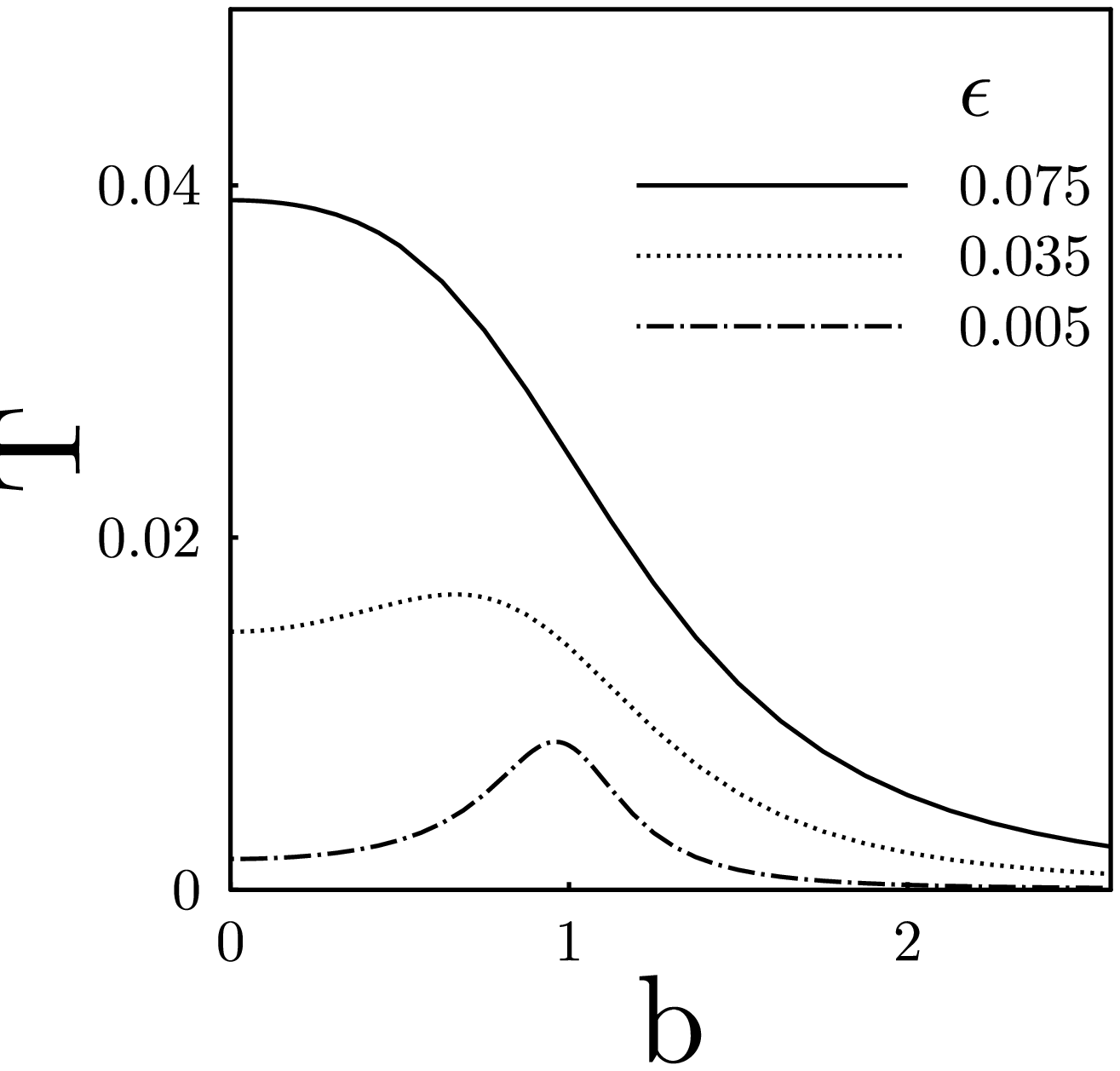}
\end{center}
\end{figure}

\vskip5cm
{\large Figure 2, F.~J.~Weiper et al.}

\newpage
\begin{figure}
\begin{center}
\leavevmode
\epsfysize=13cm
\epsffile{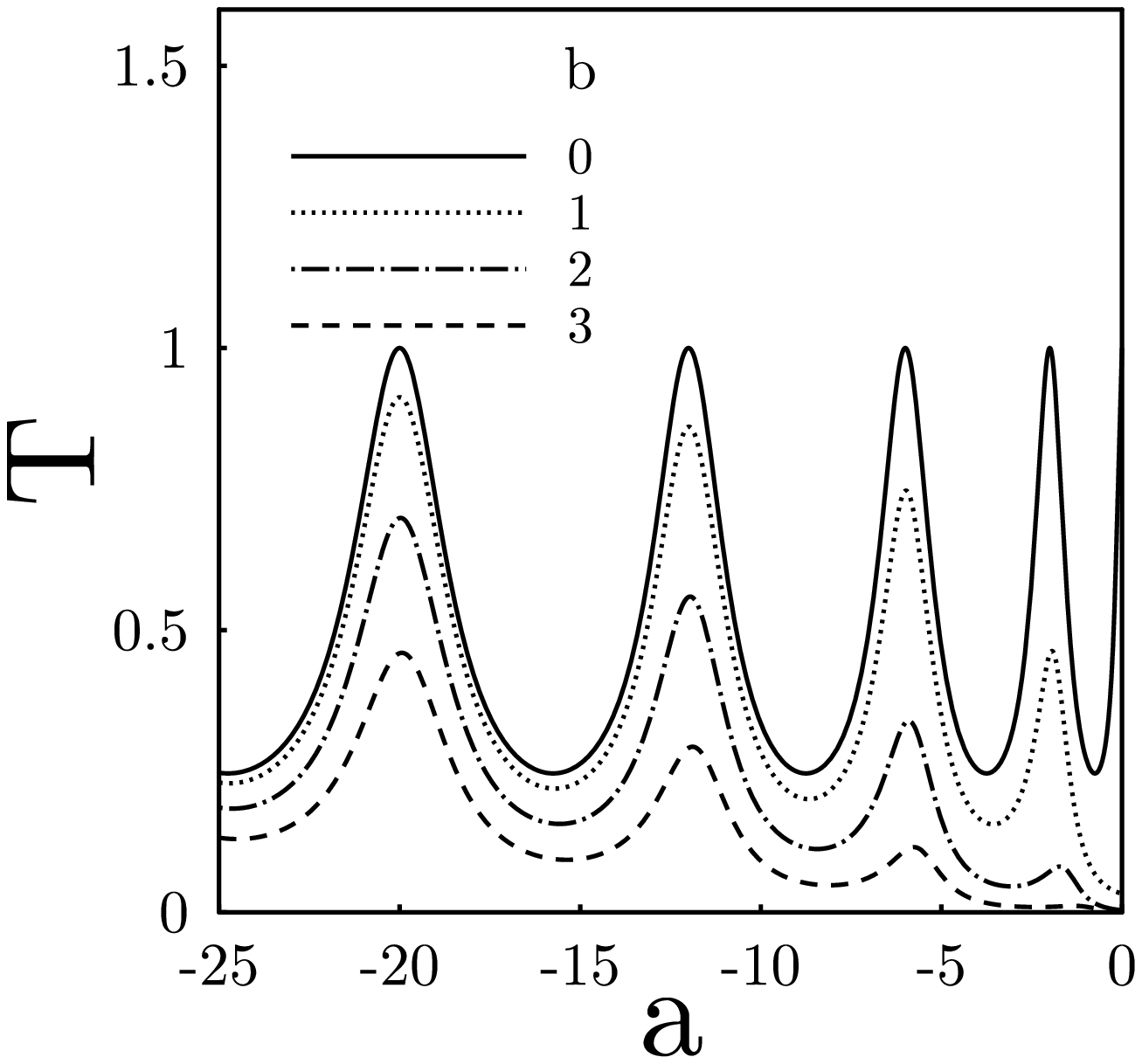}
\end{center}
\end{figure}
\vskip5cm
{\large Figure 3, F.~J.~Weiper et al.}

\newpage
\begin{figure}
\begin{center}
\vskip-0.8cm
\leavevmode
\epsfysize=13cm
\epsffile{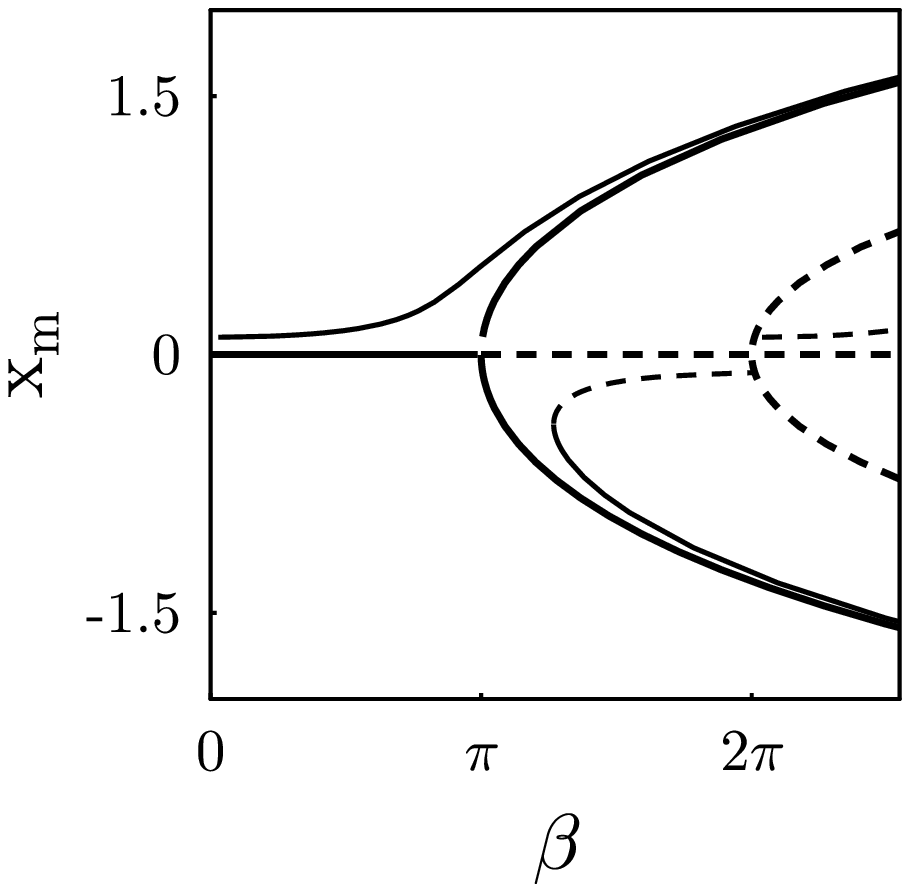}
\end{center}
\end{figure} 

\vskip5cm
{\large Figure 4, F.~J.~Weiper et al.}

\newpage

\begin{figure}
\begin{center}
\vskip-0.8cm
\leavevmode
\epsfysize=13cm
\epsffile{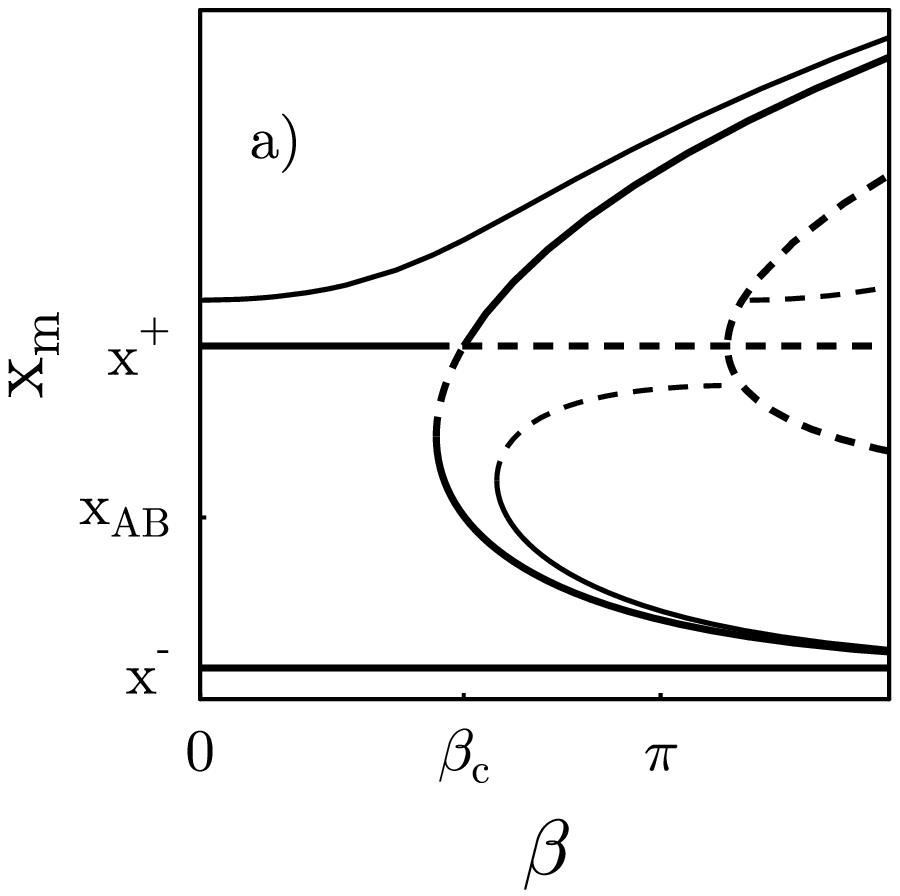}
\end{center}
\end{figure}

\vskip5cm
{\large Figure 5a, F.~J.~Weiper et al.}

\newpage

\begin{figure}
\begin{center}
\vskip-0.8cm
\leavevmode
\epsfysize=13cm
\epsffile{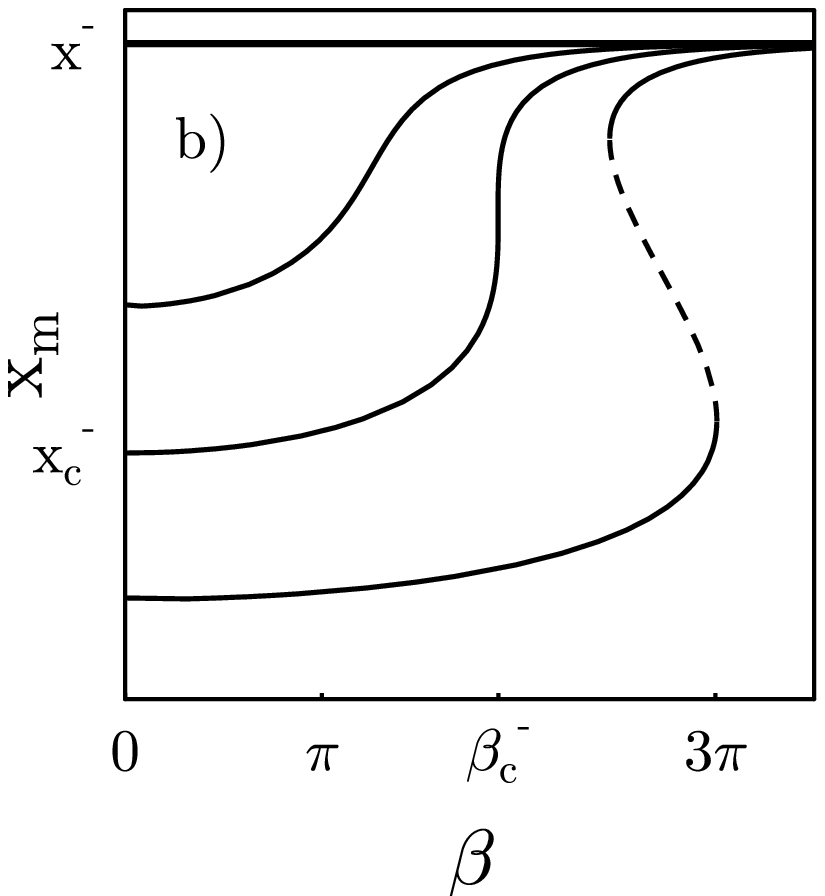}
\end{center}
\end{figure}

\vskip5cm
{\large Figure 5b, F.~J.~Weiper et al.}

\newpage
\begin{figure}
\begin{center}
\vskip-0.8cm
\leavevmode
\epsfysize=13cm
\epsffile{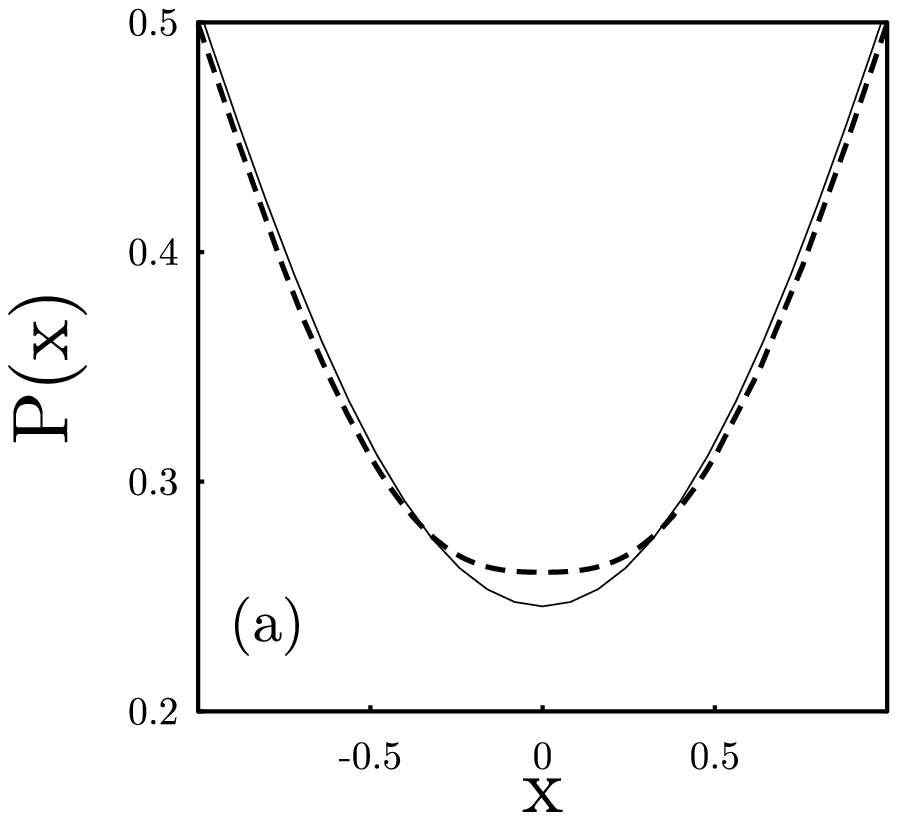}
\end{center}
\end{figure}
	
\vskip5cm
{\large Figure 6a, F.~J.~Weiper et al.}

\newpage
\begin{figure}
\begin{center}
\vskip-0.8cm
\leavevmode
\epsfysize=13cm
\epsffile{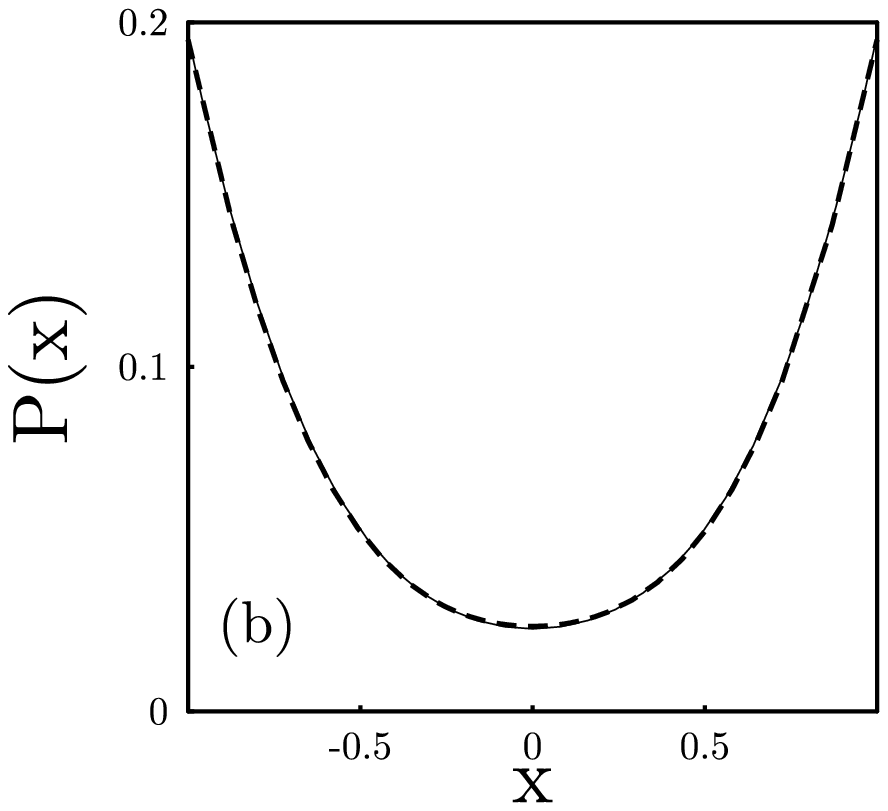}
\end{center}
\end{figure}
	
\vskip5cm
{\large Figure 6b, F.~J.~Weiper et al.}

\newpage

\begin{figure}[t]
\begin{center}
\vskip-0.8cm
\leavevmode
\epsfysize=13cm
\epsffile{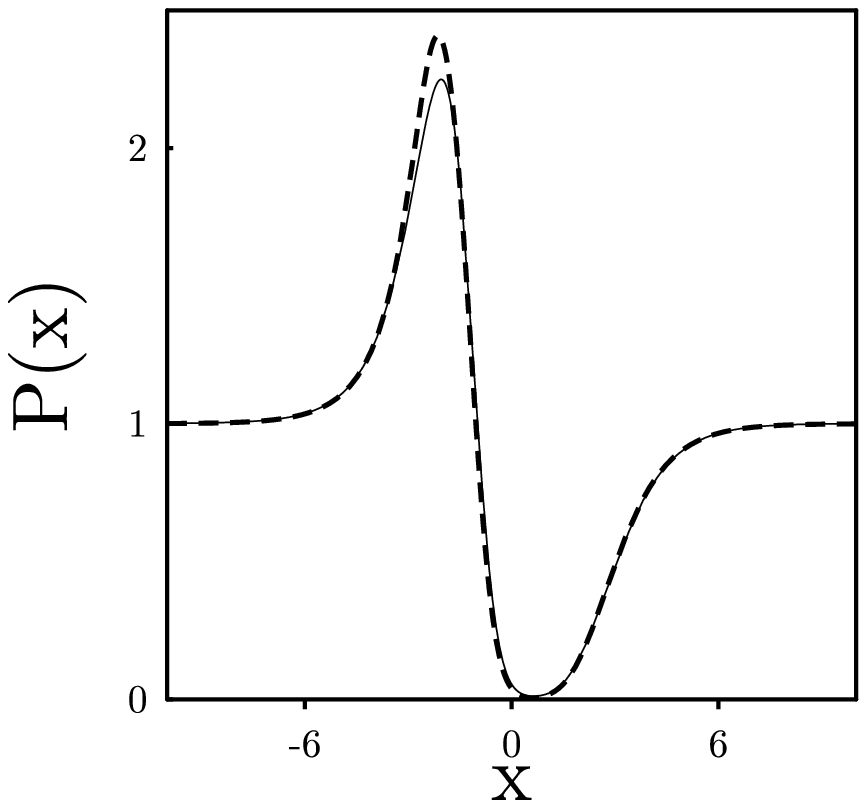}
\end{center}
\end{figure} 	

\vskip5cm
{\large Figure 7, F.~J.~Weiper et al.}

\newpage

\begin{figure}
\begin{center}
\leavevmode
\epsfysize=13cm
\epsffile{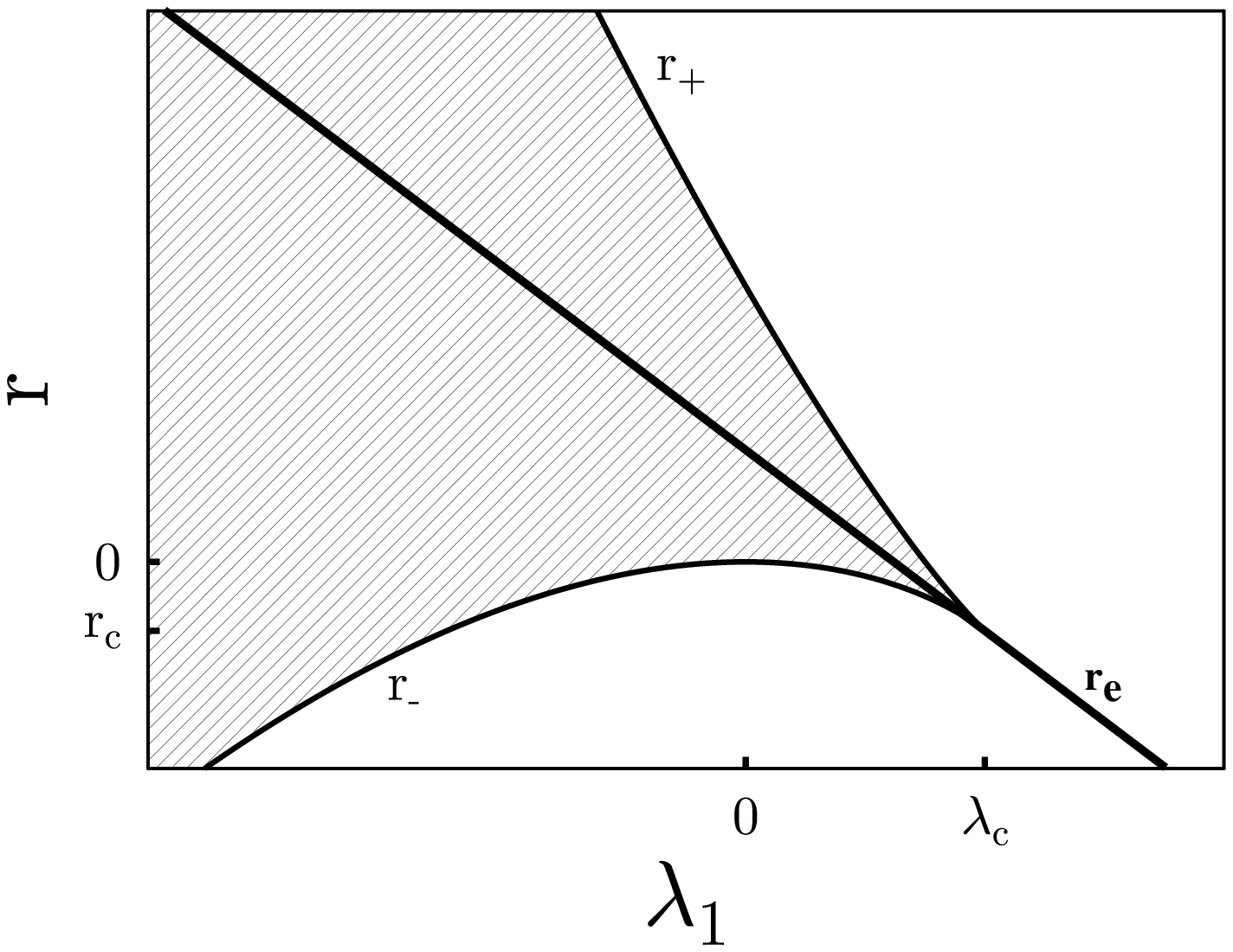}
\end{center}
\end{figure} 
 
\vskip5cm
{\large Figure 8, F.~J.~Weiper et al.}

\newpage

\begin{figure}[t]
\begin{center}
\leavevmode
\epsfysize=13cm
\epsffile{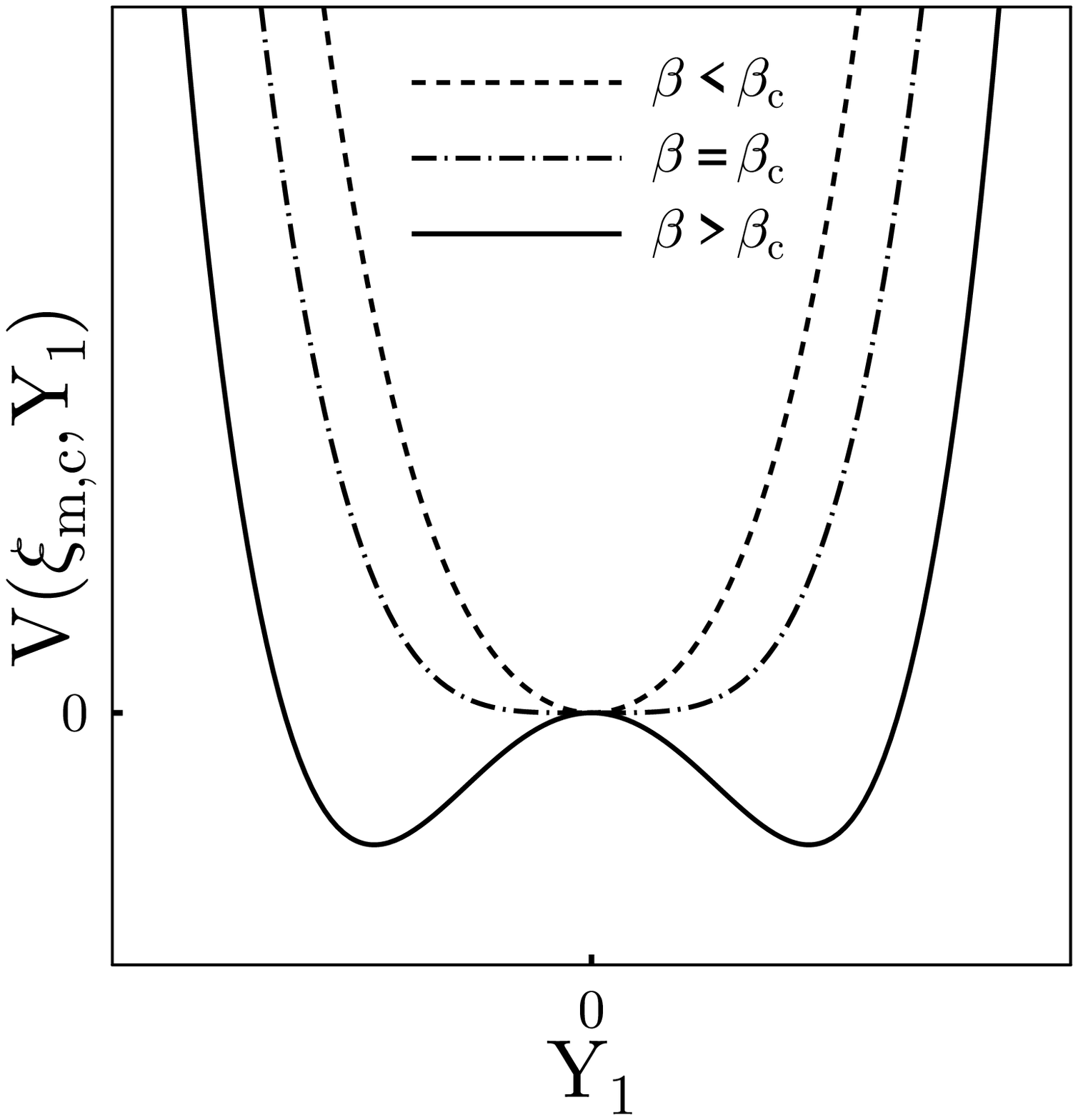}
\end{center}
\end{figure} 

\vskip5cm
{\large Figure 9, F.~J.~Weiper et al.}

\newpage
\begin{figure}[t]
\begin{center}
\vskip-0.8cm
\leavevmode
\epsfysize=13cm
\epsffile{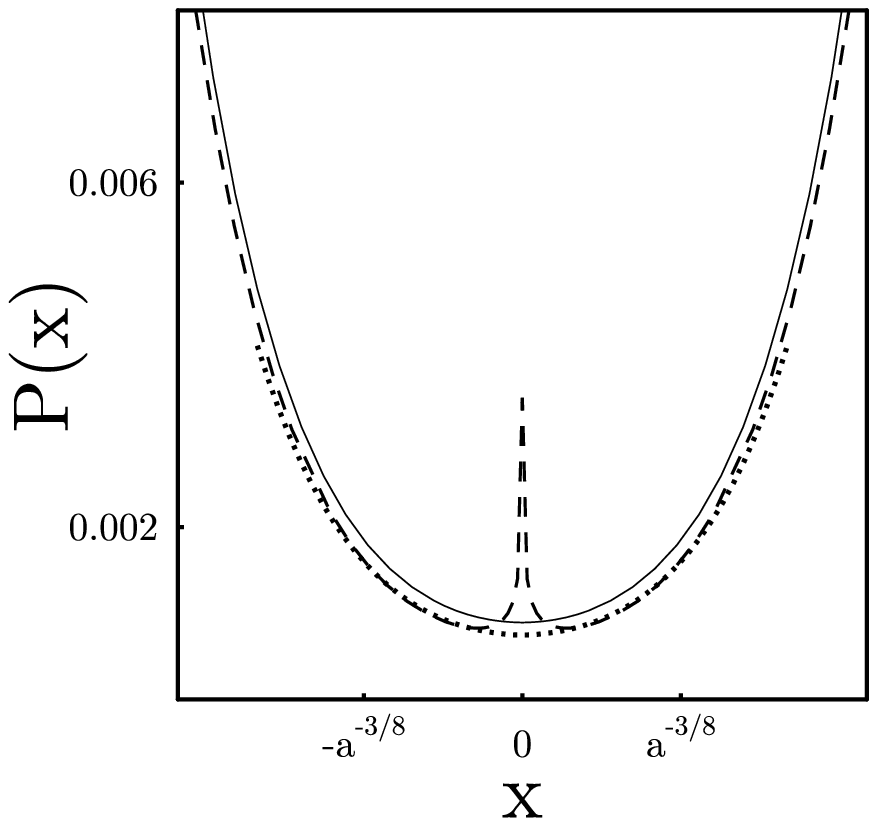}
\end{center}
\end{figure} 
	
\vskip5cm
{\large Figure 10, F.~J.~Weiper et al.}

\newpage

\begin{figure}[t]
\begin{center}
\vskip-0.8cm
\leavevmode
\epsfysize=13cm
\epsffile{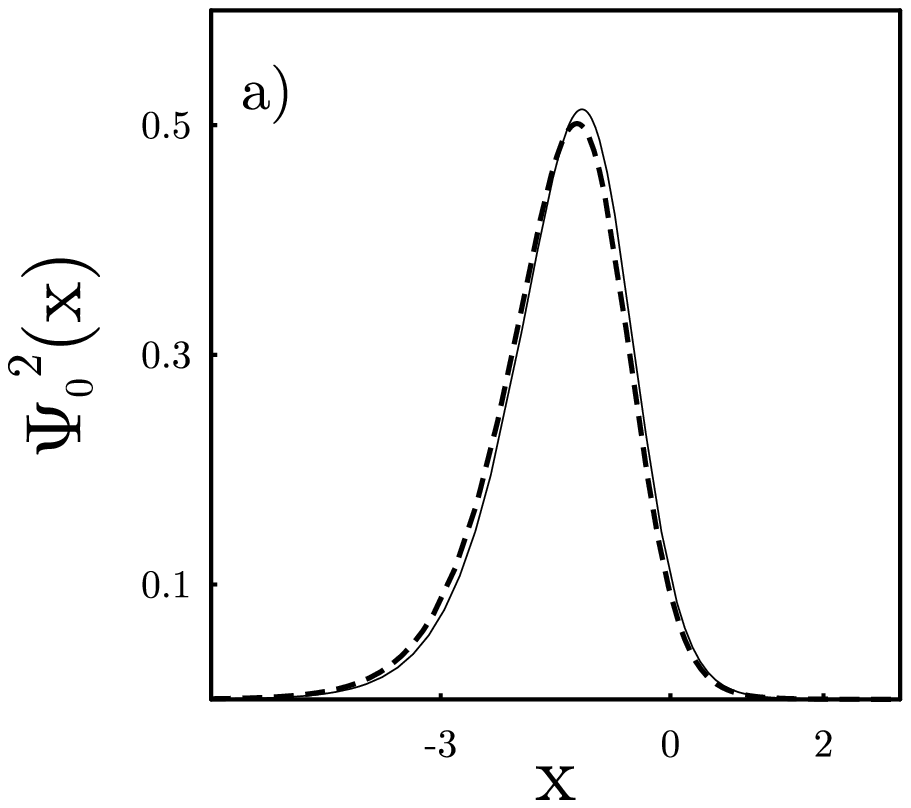}
\end{center}
\end{figure}

\vskip5cm
{\large Figure 11a, F.~J.~Weiper et al.}

\newpage

\begin{figure}[t]
\begin{center}
\vskip-0.8cm
\leavevmode
\epsfysize=13cm
\epsffile{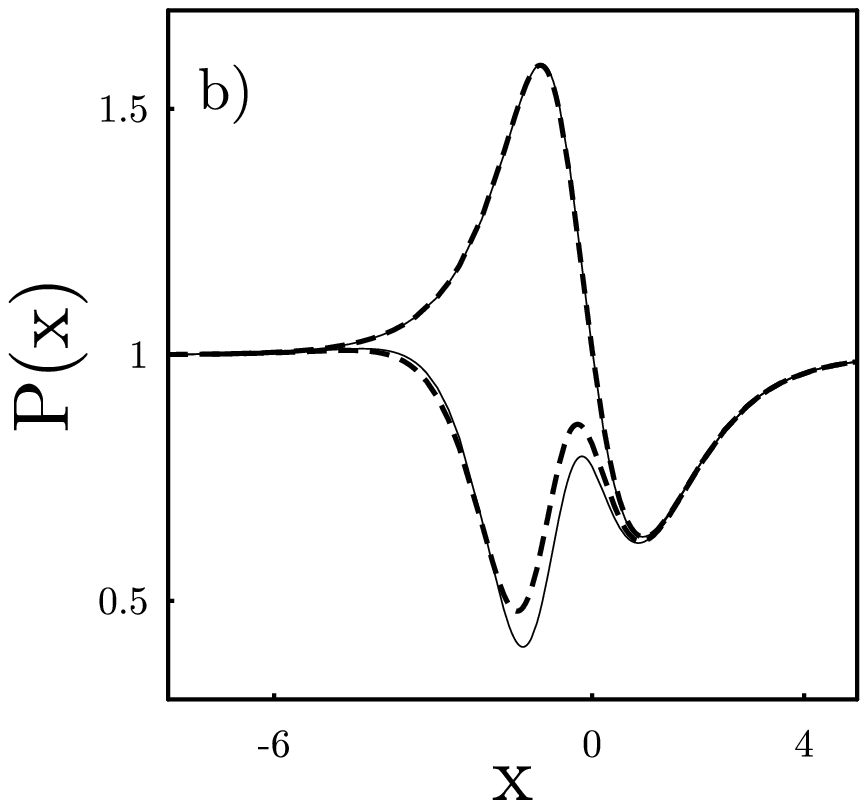}
\end{center}
\end{figure}

\vskip5cm
{\large Figure 11b, F.~J.~Weiper et al.}


\begin{thebibliography}{99}
\bibitem{footenote}
Formally, the dimensionless units introduced are obtained 
by putting $m=\hbar=L_0=1$.
\bibitem{16} 
Note that this transformation differs from
the customary ansatz $y=\tanh(x)$ for the Eckart
and Rosen--Morse potentials. The latter transformation
is not useful for potential (3) if $b\neq 0$.
\bibitem{note}
Note that we have to normalize $P_\beta (x)$ with $Z$=1 for
this procedure, i.e.~$\lim_{x\rightarrow\pm\infty}
P_\beta (x)=(2 \pi \beta)^{-1/2}$.
\end{thebibliography}
\end{document}